# Approximate $\kappa$-state Solutions to the Dirac Mobius Square-Yukawa and Mobius Square-quasi Yukawa Problems under Pseudospin and Spin Symmetry limits with Coulomb-like Tensor Interaction

Akpan N.Ikot*[1], E. Maghsoodi[2], Akaninyene D.Antia[1], S. Zarrinkamar[3] and H. Hassanabadi[2]

[1] Theoretical Physics Group, Department of Physics,University of Uyo-Nigeria.

[2] Department of Physics,Shahrood University of Technology, P.O.Box 3619995161-316, Shahrood, Iran

[3]Department of Basic Sciences, Garmsar Branch, Islamic Azad university, Garmsar, Iran

*email:ndemikot2005@yahoo.com,

**email: h.hassanabadi@shahroodu.ac.ir

**Abstract**

In this paper, we present the Dirac equation for the Mobius-square-Yukawa potentials including the tensor interaction term within the framework of pseudospin and spin symmetry limit with arbitrary spin-orbit quantum number $\kappa$. We obtained the energy eigenvalues and the corresponding wave functions using the supersymmetry method. The limiting cases of this potential model reduce to the Deng-Fan, Yukawa and Coulomb potentials, respectively.

## 1. Introduction

The concept of relativistic symmetries of the Dirac Hamiltonian discovered many years ago were recognized empirically in nuclear and hadronic spectroscopy [1]. The relativistic Dirac equation which describes the motion of spin $1/2$ particle has been successfully used in solving many physical problems in nuclear and high-energy physics [2]. Within the framework of Dirac equation, pseudospin and spin symmetries are used to study features of deformed nuclei, super-deformation and effective shell model [3]. These symmetries were originally introduced in nuclear physics and has been associated with Dirac equation as some decade ago [4]. It was shown that the exact pseudospin symmetry occurs in the Dirac equation when $\dfrac{d\Sigma(r)}{dr} = 0$, i.e. $\Sigma(r) = V(r) + S(r) = c_{ps} = const$ [5] , where $V(r), S(r)$ are repulsive and attractive scalar potentials, respectively. On the other hand, the exact spin symmetry occurs in the Dirac equation when $\dfrac{d\Delta(r)}{dr} = 0$, where $\Delta(r) = V(r) - S(r) = c_s = const$ [6]. Details of recent review of spin and pseudospin symmetries are given in Ref. [7]. These concepts, under various phenomenological potentials have been investigated by many authors using various methods such as asymptotic iteration (AIM) [8], Nikiforov-Uvarov (NU) [9], supersymmetric quantum mechanics (SUSSYQM) [10] and shape invariance(SI) [11], and exact quantization rule [12]. The pseudospin symmetry usually refers to as a quasi-degeneracy of single nucleon doublets with non-relativistic quantum number $\left(n,l, j = l + \dfrac{1}{2}\right)$ and

$\left(n-1, l+2, j=l+\frac{3}{2}\right)$, where $n, l$ and $j$ are single nucleon radial, orbital and total angular quantum numbers, respectively [4]. The total angular momentum is $j = \tilde{l} + \tilde{s}$, where $\tilde{l} = l+1$ is a pseudo-angular momentum and $\tilde{s}$ is pseudospin angular momentum [13]. Similarly, the tensor interaction term was introduced into the Dirac equation with the replacement $\vec{p} \rightarrow \vec{p} - iM\omega\beta.\hat{r}U(r)$ and a spin-orbit coupling is added to the Dirac Hamiltonian [14].The Dirac equation with different potentials in relativistic quantum mechanics with spin and pseudospin symmetry has been investigated in recent years [15-19]. Also, other potentials models investigated are given in Refs .[20-33].

The main aim of the present paper is to obtain approximate solutions of the Dirac equation with Mobius square plus Yukawa potential (MS-Y) and Mobius square plus quasi-Yukawa potential (MS-QY) potential including the Coulomb-like potential under the above mentioned symmetry limits. The paper is organized as follows. In section 2, we give a brief introduction of the supersymmetry quantum mechanics (SUSYQM) is included so that the unfamiliar reader has no problem with the mathematical tool. In section 3, Dirac equation is written for spin and pseudospin including the Coulomb interaction term. We solve the Dirac equation under these symmetries in in section 4. Few special cases are discussed in section 5. Finally, conclusion is presented in section 6.

## 2. Supersymmetry

We include this short introduction to SUSYQM to proceed on a more continues manner. In SUSUQM we normally deal with the partner Hamiltonians [10]

$$H_{\pm} = \frac{p^2}{2m} + V_{\pm}(x), \qquad (1)$$

where

$$V_{\pm}(x) = \Phi^2(x) \pm \Phi'(x). \qquad (2)$$

In the case of good SUSY, i.e. $E_0 = 0$, the ground state of the system is obtained via

$$\phi_0^-(x) = Ce^{-U}, \qquad (3)$$

where $C$ is a normalization constant and

$$U(x) = \int_{x_0}^{x} dz \Phi(z). \qquad (4)$$

Next, if the shape invariant condition

$$V_+(a_0, x) = V_-(a_1, x) + R(a_1), \qquad (5)$$

where $a_1$ is a new set of parameters uniquely determined from the old set $a_0$ via the mapping $F : a_0 \mapsto a_1 = F(a_0)$ and $R(a_1)$ does not include $x$, the higher state solutions are obtained via

$$E_n = \sum_{s=1}^{n} R(a_s), \qquad\qquad\qquad (6-a)$$

$$\phi_n^-(a_0, x) = \prod_{s=0}^{n-1} \left( \frac{A^\dagger(a_s)}{[E_n - E_s]^{1/2}} \right) \phi_0^-(a_n, x), \qquad\qquad (6-b)$$

$$\phi_0^-(a_n, x) = C \exp \left\{ -\int_0^x dz \Phi(a_n, z) \right\}, \qquad\qquad (6-c)$$

where

$$A_s^\dagger = -\frac{\partial}{\partial x} + \Phi(a_s, x). \qquad\qquad (7)$$

Therefore, this condition determines the spectrum of the bound states of the Hamiltonian

$$H_s = -\frac{\partial^2}{\partial x^2} + V_-(a_s, x) + E_s. \qquad\qquad (8)$$

and the energy eigenfunctions of

$$H_s \phi_{n-s}^-(a_s, x) = E_n \phi_{n-s}^-(a_s, x) \quad , \qquad n \geq s \qquad\qquad (9)$$

are related via [1-3]

$$\phi_{n-s}^-(a_s, x) = \frac{A^\dagger}{[E_n - E_s]^{1/2}} \phi_{n-(s+1)}^-(a_{s+1}, x). \qquad\qquad (10)$$

## 3. Dirac Equation with a Tensor Coupling

The Dirac equation for spin-$\frac{1}{2}$ particles moving in an attractive scalar potential $S(r)$, a repulsive vector potential $V(r)$ and a tensor potential $U(r)$ in the relativistic unit $(\hbar = c = 1)$ is [25]

$$\left[ \vec{\alpha}.\vec{p} + \beta(M + S(r)) - i\beta\vec{\alpha}.\hat{r}U(r) \right] \psi(r) = \left[ E - V(r) \right] \psi(r), \qquad (11),$$

where E is the relativistic energy of the system, $\vec{p} = -i\vec{\nabla}$ is the three dimensional momentum operator and M is the mass of the fermionic particle. $\vec{\alpha}, \beta$ are the $4 \times 4$ Dirac matrices given as

$$\vec{\alpha} = \begin{pmatrix} 0 & \vec{\sigma}_i \\ \vec{\sigma}_i & 0 \end{pmatrix}, \beta = \begin{pmatrix} I & 0 \\ 0 & -I \end{pmatrix}, \qquad (12)$$

where I is $2 \times 2$ unitary matrix and $\vec{\sigma}_i$ are the Pauli three-vector matrices:

$$\sigma_1 = \begin{pmatrix} 0 & 1 \\ 1 & 0 \end{pmatrix}, \sigma_2 = \begin{pmatrix} 0 & -i \\ i & 0 \end{pmatrix}, \sigma_3 = \begin{pmatrix} 1 & 0 \\ 0 & -1 \end{pmatrix}. \qquad (13)$$

The eigenvalues of the spin-orbit coupling operator are $\kappa = \left(j+\dfrac{1}{2}\right) \succ 0, \kappa = -\left(j+\dfrac{1}{2}\right) \prec 0$ for unaligned $j = l-\dfrac{1}{2}$ and the aligned spin $j = l+\dfrac{1}{2}$ respectively. The set $\left(H, K, J^2, J_z\right)$ forms a complete set of conserved quantities. Thus, we can write the spinors as [26],

$$\psi_{n\kappa}(r) = \frac{1}{r}\begin{pmatrix} F_{n\kappa}(r)Y^l_{jm}(\theta,\varphi) \\ iG_{n\kappa}(r)Y^{\tilde{l}}_{jm}(\theta,\varphi) \end{pmatrix} \qquad (14)$$

where $F_{n\kappa}(r), G_{n\kappa}(r)$ represent the upper and lower components of the Dirac spinors. $Y^l_{jm}(\theta,\varphi), Y^{\tilde{l}}_{jm}(\theta,\varphi)$ are the spin and pseudospin spherical harmonics and $m$ is the projection on the z-axis. With other known identities [27],

$$\left(\vec{\sigma}.\vec{A}\right)\left(\vec{\sigma}.\vec{B}\right) = \vec{A}.\vec{B} + i\,\vec{\sigma}.\left(\vec{A}x\vec{B}\right),$$

$$\vec{\sigma}.\vec{p} = \vec{\sigma}.\hat{r}\left(\hat{r}.\vec{p} + i\,\frac{\vec{\sigma}.\vec{L}}{r}\right) \qquad (15)$$

as well as

$$\left(\vec{\sigma}.\vec{L}\right)Y^{\tilde{l}}_{jm}(\theta,\varphi) = (\kappa-1)Y^{\tilde{l}}_{jm}(\theta,\varphi)$$

$$\left(\vec{\sigma}.\vec{L}\right)Y^l_{jm}(\theta,\varphi) = -(\kappa-1)Y^l_{jm}(\theta,\varphi)$$

$$\left(\vec{\sigma}.\hat{r}\right)Y^l_{jm}(\theta,\varphi) = -Y^{\tilde{l}}_{jm}(\theta,\varphi) \qquad (16)$$

$$\left(\vec{\sigma}.\hat{r}\right)Y^{\tilde{l}}_{jm}(\theta,\varphi) = -Y^l_{jm}(\theta,\varphi)$$

we find the following two coupled first-order Dirac equation [27],

$$\left(\frac{d}{dr} + \frac{\kappa}{r} - U(r)\right)F_{n\kappa}(r) = \left(M + E_{n\kappa} - \Delta(r)\right)G_{n\kappa}(r), \quad (17)$$

$$\left(\frac{d}{dr} - \frac{\kappa}{r} + U(r)\right)G_{n\kappa}(r) = \left(M - E_{n\kappa} + \Sigma(r)\right)F_{n\kappa}(r), \qquad (18)$$

where,

$$\Delta(r) = V(r) - S(r) \qquad (19)$$

$$\Sigma(r) = V(r) + S(r) \qquad (20)$$

Eliminating $F_{n\kappa}(r)$ and $G_{n\kappa}(r)$ in Eqs.(17) and (18), we obtain the second-order Schrödinger-like equation as,

$$\left\{ \begin{array}{l} \dfrac{d^2}{dr^2} - \dfrac{\kappa(\kappa+1)}{r^2} + \dfrac{2\kappa U(r)}{r} - \dfrac{dU(r)}{dr} - U^2(r) - \left(M + E_{n\kappa} - \Delta(r)\right)\left(M - E_{n\kappa} + \Sigma(r)\right) \\ + \dfrac{\dfrac{d\Delta(r)}{dr}\left(\dfrac{d}{dr} + \dfrac{\kappa}{r} - U(r)\right)}{\left(M + E_{n\kappa} - \Delta(r)\right)} \end{array} \right\} F_{n\kappa}(r) = 0, \quad (21)$$

$$\begin{cases} \dfrac{d^2}{dr^2} - \dfrac{\kappa(\kappa-1)}{r^2} + \dfrac{2\kappa U(r)}{r} + \dfrac{dU(r)}{dr} - U^2(r) - \big(M + E_{n\kappa} - \Delta(r)\big)\big(M - E_{n\kappa} + \Sigma(r)\big) \\[2mm] \quad + \dfrac{\dfrac{d\Sigma(r)}{dr}\left(\dfrac{d}{dr} - \dfrac{\kappa}{r} + U(r)\right)}{\big(M + E_{n\kappa} - \Sigma(r)\big)} \end{cases} G_{n\kappa}(r) = 0, \quad (22)$$

where $\kappa(\kappa-1) = \tilde{l}(\tilde{l}+1), \kappa(\kappa+1) = l(l+1)$. The radial wave functions are required to satisfy the necessary conditions that is $F_{n\kappa}$ and $G_{n\kappa}$ vanish at the origin and the infinity. At this stage, we take $\Delta(r)$ or $\Sigma(r)$ as the MS-Y potential. Eqs. (21) and (22) can be exactly solved for $\kappa = 0, -1$ and $\kappa = 0, 1$, respectively, as the spin-orbit centrifugal term vanishes.

## 4. Solution of the first choice

In this section, we are going to solve the Dirac equation with MS-Y potential and tensor potential by using the SUSYQM.

### 4.1. Pseudospin Symmetry Limit for the first choice

The exact pseudospin symmetry is proved by Meng et al. [28]. It occurs in Dirac equation when $\dfrac{d\Sigma(r)}{dr} = 0$ or $\Sigma(r) = C_{ps} = const$ [5]. In this limit, we take $\Delta(r)$ as the MS-Y potential and a Coulomb-like potential [29] for the tensor potential added,

$$\Delta(r) = V_0\left(\frac{A^{ps} + B^{ps}e^{-\alpha r}}{C^{ps} + D^{ps}e^{-\alpha r}}\right)^2 - V_1\frac{e^{-\alpha r}}{r}, \qquad (23)$$

$$U(r) = -\frac{H}{r}, H = \frac{z_a z_b e^2}{4\pi\varepsilon_0}, r \geq R_e \qquad (24)$$

where $V_0, V_1, A, B, C, D$ and $H$ are constant coefficients, $R_e = 7.78\,fm$ is the Coulomb radius, $z_a$ and $z_b$ denotes the charges of the projectile a and target nuclei b, respectively [29]. Since Dirac equation with the MS-Y potential has no exact solution, we use an approximation for the centrifugal term as shown in fig.(1),

$$\frac{1}{r^2} = \lim_{\alpha \to 0}\left[\frac{C^{ps2}\alpha^{ps2}}{\left(C^{ps} + D^{ps}e^{-\alpha r}\right)^2}\right], \quad (25)$$

in Eq. (22) yields

$$\left\{-\frac{d^2}{dr^2} + \frac{1}{(1 + \frac{D^{ps}}{C^{ps}}e^{-\alpha r})^2}\left\{\left[\frac{-MV_0 B^{ps2}}{C^{ps2}} + \frac{V_0 E_{n\kappa}^{ps} B^{ps2}}{C^{ps2}} - \frac{V_0 C_{ps} B^{ps2}}{C^{ps2}} + \frac{MV_1\alpha D^{ps}}{C^{ps}} - \frac{E_{n\kappa}^{ps} V_1\alpha D^{ps}}{C^{ps}} + \frac{C_{ps}V_1\alpha D^{ps}}{C^{ps}}\right]e^{-2\alpha r}\right.\right.$$

$$+\left[-\frac{2A^{ps}B^{ps}V_0 M}{C^{ps2}} + \frac{2A^{ps}B^{ps}V_0 E_{n\kappa}^{ps}}{C^{ps2}} - \frac{2A^{ps}B^{ps}V_0 C_{ps}}{C^{ps2}} + MV_1\alpha - E_{n\kappa}^{ps}V_1\alpha + C_{ps}V_1\alpha\right]e^{-\alpha r} + (\kappa + H)(\kappa + H - 1)\alpha^2$$

$$\left.-\frac{MV_0 A^{ps2}}{C^{ps2}} + \frac{V_0 E_{n\kappa}^{ps} A^{ps2}}{C^{ps2}} - \frac{V_0 C_{ps} A^{ps2}}{C^{ps2}}\right\}G_{n\kappa}^{ps}(r) = \left\{-M^2 - MC_{ps} + (E_{n\kappa}^{ps})^2 - E_{n\kappa}^{ps}C_{ps}\right\}G_{n\kappa}^{ps}(r), \quad (26)$$

where $\kappa = -\tilde{\ell}$ and $\kappa = \tilde{\ell}+1$ for $\kappa < 0$ and $\kappa > 0$.

## 4.2. Solution Pseudospin Symmetry Limit for the first choice

In the previous section, we obtained a Schrödinger-like equation of the form

$$-\frac{d^2 G_{n\kappa}^{ps}(r)}{dr^2} + V_{eff}(r)G_{n\kappa}^{ps}(r) = \tilde{E}_{n\kappa}^{ps}G_{n\kappa}^{ps}(r), \qquad (27)$$

with the effective potential being

$$V_{eff} = \frac{\eta^{ps}e^{-2\alpha r} + \zeta^{ps}e^{-\alpha r} + \sigma^{ps}}{(1 + \frac{D^{ps}}{C^{ps}}e^{-\alpha r})^2}, \qquad (28)$$

where

$$\eta^{ps} = \frac{-MV_0 B^{ps2}}{C^{ps2}} + \frac{V_0 E_{n\kappa}^{ps}B^{ps2}}{C^{ps2}} - \frac{V_0 C_{ps}B^{ps2}}{C^{ps2}} + \frac{MV_1\alpha D^{ps}}{C^{ps}} - \frac{E_{n\kappa}^{ps}V_1\alpha D^{ps}}{C^{ps}} + \frac{C_{ps}V_1\alpha D^{ps}}{C^{ps}},$$

$$\zeta^{ps} = -\frac{2A^{ps}B^{ps}V_0 M}{C^{ps2}} + \frac{2A^{ps}B^{ps}V_0 E_{n\kappa}^{ps}}{C^{ps2}} - \frac{2A^{ps}B^{ps}V_0 C_{ps}}{C^{ps2}} + MV_1\alpha - E_{n\kappa}^{ps}V_1\alpha + C_{ps}V_1\alpha,$$

$$\sigma^{ps} = (\kappa + H)(\kappa + H - 1)\alpha^2 - \frac{MV_0 A^{ps2}}{C^{ps2}} + \frac{V_0 E_{n\kappa}^{ps}A^{ps2}}{C^{ps2}} - \frac{V_0 C_{ps}A^{ps2}}{C^{ps2}}. \qquad (29)$$

The corresponding effective energy is given by

$$\tilde{E}_{n\kappa}^{ps} = -M^2 - MC_{ps} + (E_{n\kappa}^{ps})^2 - E_{n\kappa}^{ps}C_{ps}, \qquad (30)$$

In SUSYQM formalism, the ground-state wave function for the lower component is given as

$$G_{0\kappa}^{ps}(r) = \exp\left(-\int \phi(r)\,dr\right), \qquad (31)$$

Thus, we are dealing with the Riccati equation

$$\phi^2 - \phi' = V_{eff} - \tilde{E}_{0\kappa}^{ps}, \qquad (32)$$

for which we propose a superpotential of the form

$$\phi(r) = \frac{f^{ps}e^{-\alpha r}}{(1 + \frac{D^{ps}}{C^{ps}}e^{-\alpha r})} + g^{ps}. \qquad (33)$$

Therefore, the exact parameters of our study are obtained via

$$\frac{(f^{ps})^2 e^{-2\alpha r}}{(1+\frac{D^{ps}}{C^{ps}}e^{-\alpha r})^2} + (g^{ps})^2 + \frac{2 f^{ps} g^{ps} e^{-\alpha r}}{(1+\frac{D^{ps}}{C^{ps}}e^{-\alpha r})} + \frac{f^{ps}\alpha e^{-\alpha r}}{(1+\frac{D^{ps}}{C^{ps}}e^{-\alpha r})^2} =$$

$$\frac{\eta^{ps}e^{-2\alpha r} + \zeta^{ps}e^{-\alpha r} + \sigma^{ps}}{(1+\frac{D^{ps}}{C^{ps}}e^{-\alpha r})^2} - \tilde{E}_{0\kappa}^{ps}, \qquad (34)$$

or, more explicitly

$$\tilde{E}_{0\kappa}^{ps} = \sigma^{ps} - (g^{ps})^2, \qquad\qquad (35-a)$$

$$f^{ps} = \frac{\alpha D^{ps}}{2C^{ps}} - \frac{1}{2C^{ps}}\sqrt{\alpha^2 (D^{ps})^2 + 4\sigma^{ps}(D^{ps})^2 + 4\eta^{ps}(C^{ps})^2 - 4\zeta^{ps}C^{ps}D^{ps}}, \qquad (35-b)$$

$$g^{ps} = \frac{-(f^{ps})^2 (C^{ps})^2 - \sigma^{ps}(D^{ps})^2 + \eta^{ps}(C^{ps})^2}{2C^{ps}D^{ps}f^{ps}}. \qquad (35-c)$$

After constructing the partner Hamiltonians

$$V_{eff+}(r) = \phi^2 + \frac{d\phi}{dr}$$

$$= \frac{-f^{ps}\frac{C^{ps}}{D^{ps}}[f^{ps} + \alpha\frac{D^{ps}}{C^{ps}}]e^{-\alpha r}}{(1+\frac{D^{ps}}{C^{ps}}e^{-\alpha r})^2} + \frac{[\frac{-\sigma^{ps}(D^{ps})^2 + \eta^{ps}(C^{ps})^2}{C^{ps}D^{ps}}]e^{-\alpha r}}{(1+\frac{D^{ps}}{C^{ps}}e^{-\alpha r})}$$

$$+ (\frac{-(f^{ps})^2 (C^{ps})^2 - \sigma^{ps}(D^{ps})^2 + \eta^{ps}(C^{ps})^2}{2C^{ps}D^{ps}f^{ps}})^2, \qquad (36-a)$$

$$V_{eff-}(r) = \phi^2 - \frac{d\phi}{dr}$$

$$\frac{-f^{ps}\frac{C^{ps}}{D^{ps}}[f^{ps} - \alpha\frac{D^{ps}}{C^{ps}}]e^{-\alpha r}}{(1+\frac{D^{ps}}{C^{ps}}e^{-\alpha r})^2} + \frac{[\frac{-\sigma^{ps}(D^{ps})^2 + \eta^{ps}(C^{ps})^2}{C^{ps}D^{ps}}]e^{-\alpha r}}{(1+\frac{D^{ps}}{C^{ps}}e^{-\alpha r})}$$

$$+ (\frac{-(f^{ps})^2 (C^{ps})^2 - \sigma^{ps}(D^{ps})^2 + \eta^{ps}(C^{ps})^2}{2C^{ps}D^{ps}f^{ps}})^2, \qquad (36-b)$$

where $a_0 = f^{ps}$ and $a_i$ is a function of $a_0$, i.e., $a_1 = f(a_0) = a_0 + \alpha\frac{D^{ps}}{C^{ps}}$. Consequently,

$a_n = f(a_0) = a_0 + n\alpha\frac{D^{ps}}{C^{ps}}$. We see that the shape invariance holds via a mapping of the form

$f^{ps} \to f^{ps} + \alpha\frac{D^{ps}}{C^{ps}}$ . From Eq. (5), we have [29, 30]

$$R(a_1) = (\frac{-a_0^2(C^{ps})^2 - \sigma^{ps}(D^{ps})^2 + \eta^{ps}(C^{ps})^2}{2C^{ps}D^{ps}a_0})^2 - (\frac{-a_1^2(C^{ps})^2 - \sigma^{ps}(D^{ps})^2 + \eta^{ps}(C^{ps})^2}{2C^{ps}D^{ps}a_1})^2,$$

$$R(a_2) = (\frac{-a_1^2(C^{ps})^2 - \sigma^{ps}(D^{ps})^2 + \eta^{ps}(C^{ps})^2}{2C^{ps}D^{ps}a_1})^2 - (\frac{-a_2^2(C^{ps})^2 - \sigma^{ps}(D^{ps})^2 + \eta^{ps}(C^{ps})^2}{2C^{ps}D^{ps}a_2})^2,$$

$$R(a_3) = (\frac{-a_2^2(C^{ps})^2 - \sigma^{ps}(D^{ps})^2 + \eta^{ps}(C^{ps})^2}{2C^{ps}D^{ps}a_2})^2 - (\frac{-a_3^2(C^{ps})^2 - \sigma^{ps}(D^{ps})^2 + \eta^{ps}(C^{ps})^2}{2C^{ps}D^{ps}a_3})^2,$$

.

.

$$R(a_n) = (\frac{-a_{n-1}^2(C^{ps})^2 - \sigma^{ps}(D^{ps})^2 + \eta^{ps}(C^{ps})^2}{2C^{ps}D^{ps}a_{n-1}})^2 - (\frac{-a_n^2(C^{ps})^2 - \sigma^{ps}(D^{ps})^2 + \eta^{ps}(C^{ps})^2}{2C^{ps}D^{ps}a_n})^2, \quad (37)$$

$$\tilde{E}_{0\kappa}^- = 0 \qquad (38)$$

Therefore, the from Eq. (6-a) the eigenvalues can be found as

$$\tilde{E}_{n\kappa}^{ps-} = \sum_{k=1}^n R(a_\kappa) = (\frac{-a_0^2(C^{ps})^2 - \sigma^{ps}(D^{ps})^2 + \eta^{ps}(C^{ps})^2}{2C^{ps}D^{ps}a_0})^2$$

$$- (\frac{-a_n^2(C^{ps})^2 - \sigma^{ps}(D^{ps})^2 + \eta^{ps}(C^{ps})^2}{2C^{ps}D^{ps}a_n})^2, \qquad (39-a)$$

$$\tilde{E}_{n\kappa}^{ps} = \tilde{E}_{n\kappa}^{ps-} + \tilde{E}_{0\kappa}^{ps} = -(\frac{-a_n^2(C^{ps})^2 - \sigma^{ps}(D^{ps})^2 + \eta^{ps}(C^{ps})^2}{2C^{ps}D^{ps}a_n})^2 + \sigma^{ps}, \qquad (39-b)$$

This completely determines the energy of the pseudospin symmetry limit.

With the aid of Eqs.(29) and (35-a)-(35-c), we obtain the energy eigenvalues for the MS-Y potential model for the pseudospin symmetry limit for any spin-orbit quantum number as

$$-M^2 - MC_{ps} + (E_{n\kappa}^{ps})^2 - E_{n\kappa}^{ps}C_{ps} + \frac{1}{\left\{2C^{ps}D^{ps}\left[\left(\frac{\alpha D^{ps}}{2C^{ps}} \pm \frac{1}{2C^{ps}}\left(\sqrt{\begin{array}{c}\alpha^2(D^{ps})^2 + 4\sigma^{ps}(D^{ps})^2 \\ + 4\eta^{ps}(C^{ps})^2 - 4\zeta^{ps}C^{ps}D^{ps}\end{array}}\right)\right) + n\alpha\frac{D^{ps}}{C^{ps}}\right]\right\}^2}$$

$$\times\left\{-\left[\left(\frac{\alpha D^{ps}}{2C^{ps}} \pm \frac{1}{2C^{ps}}\left(\sqrt{\begin{array}{c}\alpha^2(D^{ps})^2 + 4\sigma^{ps}(D^{ps})^2 \\ + 4\eta^{ps}(C^{ps})^2 - 4\zeta^{ps}C^{ps}D^{ps}\end{array}}\right)\right) + n\alpha\frac{D^{ps}}{C^{ps}}\right]^2(C^{ps})^2 - \sigma^{ps}(D^{ps})^2 + \eta^{ps}(C^{ps})^2\right\} - \sigma^{ps} = 0. \quad (40)$$

Thus, the lower component of the wave function as,

$$G_{n\kappa}(r) = N_{n\kappa}\left(-\frac{D^{ps}}{C^{ps}}e^{-\alpha r}\right)^{\sqrt{\omega_3}}\left(1 + \frac{D^{ps}}{C^{ps}}e^{-\alpha r}\right)^{\frac{1}{2} + \sqrt{\omega_1 - \omega_2 + \omega_3 + \frac{1}{4}}}P_n^{(2\sqrt{\omega_3},\, 2\sqrt{\omega_1 - \omega_2 + \omega_3 + \frac{1}{4}})}(1 + \frac{2D^{ps}}{C^{ps}}e^{-\alpha r}), \quad (41)$$

where

$$\omega_1 = -\frac{MV_0B^{ps2}}{\alpha^2D^{ps2}} + \frac{V_0E_{n\kappa}^{ps}B^{ps2}}{\alpha^2D^{ps2}} - \frac{V_0C_{ps}B^{ps2}}{\alpha^2D^{ps2}} + \frac{MV_1C^{ps}}{\alpha D^{ps}} - \frac{E_{n\kappa}^{ps}V_1C^{ps}}{\alpha D^{ps}} + \frac{C_{ps}V_1C^{ps}}{\alpha D^{ps}} + \frac{M^2}{\alpha^2} + \frac{MC_{ps}}{\alpha^2} - \frac{(E_{n\kappa}^{ps})^2}{\alpha^2} + \frac{E_{n\kappa}^{ps}C_{ps}}{\alpha^2},$$

$$\omega_2 = -\frac{2A^{ps}B^{ps}V_0M}{\alpha^2D^{ps}C^{ps}} + \frac{2A^{ps}B^{ps}V_0E_{n\kappa}^{ps}}{\alpha^2D^{ps}C^{ps}} - \frac{2A^{ps}B^{ps}V_0C_{ps}}{\alpha^2D^{ps}C^{ps}} + \frac{MV_1C^{ps}}{\alpha D^{ps}} - \frac{E_{n\kappa}^{ps}V_1C^{ps}}{\alpha D^{ps}} + \frac{C_{ps}V_1C^{ps}}{\alpha D^{ps}} + \frac{2M^2}{\alpha^2} + \frac{2MC_{ps}}{\alpha^2}$$

$$- \frac{2(E_{n\kappa}^{ps})^2}{\alpha^2} + \frac{2E_{n\kappa}^{ps}C_{ps}}{\alpha^2},$$

$$\omega_3 = (\kappa+H)(\kappa+H-1) - \frac{MV_0A^{ps2}}{\alpha^2C^{ps2}} + \frac{V_0E_{n\kappa}^{ps}A^{ps2}}{\alpha^2C^{ps2}} - \frac{V_0C_{ps}A^{ps2}}{\alpha^2C^{ps2}} + \frac{M^2}{\alpha^2} + \frac{MC_{ps}}{\alpha^2} - \frac{(E_{n\kappa}^{ps})^2}{\alpha^2} + \frac{E_{n\kappa}^{ps}C_{ps}}{\alpha^2}. \quad (42)$$

and $N_{n\kappa}$ is the normalization constant. The upper spinor component of the Dirac equation can be calculated as ,

$$F_{n\kappa}(r) = \frac{1}{M - E_{n\kappa} + C_{ps}}\left(\frac{d}{dr} - \frac{\kappa}{r} + U(r)\right)G_{n\kappa}(r), \quad (43)$$

where $E_{n\kappa} \neq M + C_{ps}$ and when $C_{ps} = 0$ (exact pseudospin symmetry) is obtained which means that only negative energy solutions are possible.

### 4.3. Spin Symmetry Limit for the first Choice

In the spin symmetry limit $\frac{d\Delta(r)}{dr} = 0$ or $\Delta(r) = C_s = const.$[17, 18]. As the previous section, we consider [19]

$$\Sigma(r) = V_0\left(\frac{A^s + B^se^{-\alpha r}}{C^s + D^se^{-\alpha r}}\right)^2 - V_1\frac{e^{-\alpha r}}{r}, \quad (44)$$

Substitution of the latter in Eq. (9) gives

$$\{-\frac{d^2}{dr^2} + \frac{1}{(1 + \frac{D^s}{C^s}e^{-\alpha r})^2}\{[\frac{MV_0B^{s2}}{C^{s2}} + \frac{V_0E_{n\kappa}^sB^{s2}}{C^{s2}} - \frac{V_0C_sB^{s2}}{C^{s2}} - \frac{MV_1\alpha D^s}{C^s} - \frac{E_{n\kappa}^sV_1\alpha D^s}{C^s} + \frac{C_sV_1\alpha D^s}{C^s}]e^{-2\alpha r}$$

$$+ [\frac{2A^sB^sV_0M}{C^{s2}} + \frac{2A^sB^sV_0E_{n\kappa}^s}{C^{s2}} - \frac{2A^sB^sV_0C_s}{C^{s2}} - MV_1\alpha - E_{n\kappa}^sV_1\alpha + C_sV_1\alpha]e^{-\alpha r} + (\kappa+H+1)(\kappa+H)\alpha^2$$

$$+ \frac{MV_0A^{s2}}{C^{s2}} + \frac{V_0E_{n\kappa}^sA^{s2}}{C^{s2}} - \frac{V_0C_sA^{s2}}{C^{s2}}\}F_{n\kappa}^s(r) = \{-M^2 + MC_s + (E_{n\kappa}^s)^2 - E_{n\kappa}^sC_s\}F_{n\kappa}^s(r), \quad (45)$$

where $\kappa = \ell$ and $\kappa = -\ell - 1$ for $\kappa < 0$ and $\kappa > 0$ .

### 4.4. Solution of the Spin Symmetry Limit for the First Choice

In this case,

$$-\frac{d^2F_{n\kappa}^s(r)}{dr^2} + V_{eff}(r)F_{n\kappa}^s(r) = \tilde{E}_{n\kappa}^s F_{n\kappa}^s(r), \quad (46)$$

with

$$V_{eff} = \frac{\eta^s e^{-2\alpha r} + \zeta^s e^{-\alpha r} + \sigma^s}{(1 + \frac{D^s}{C^s} e^{-\alpha r})^2}, \qquad (47)$$

where

$$\eta^s = \frac{MV_0 B^{s2}}{C^{s2}} + \frac{V_0 E_{n\kappa}^s B^{s2}}{C^{s2}} - \frac{V_0 C_s B^{s2}}{C^{s2}} - \frac{MV_1 \alpha D^s}{C^s} - \frac{E_{n\kappa}^s V_1 \alpha D^s}{C^s} + \frac{C_s V_1 \alpha D^s}{C^s},$$

$$\zeta^s = \frac{2A^s B^s V_0 M}{C^{s2}} + \frac{2A^s B^s V_0 E_{n\kappa}^s}{C^{s2}} - \frac{2A^s B^s V_0 C_s}{C^{s2}} - MV_1 \alpha - E_{n\kappa}^s V_1 \alpha + C_s V_1 \alpha,$$

$$\sigma^s = (\kappa + H + 1)(\kappa + H)\alpha^2 + \frac{MV_0 A^{s2}}{C^{s2}} + \frac{V_0 E_{n\kappa}^s A^{s2}}{C^{s2}} - \frac{V_0 C_s A^{s2}}{C^{s2}}. \qquad (48)$$

and

$$\tilde{E}_{n\kappa}^s = -M^2 + MC_s + (E_{n\kappa}^s)^2 - E_{n\kappa}^s C_s, \qquad (49)$$

By considering the same way of solving equation (45), the energy equation for the MS-Y potential in the presence of tensor interaction in view of the spin symmetry limit is obtained as follows

$$-M^2 + MC_s + (E_{n\kappa}^s)^2 - E_{n\kappa}^s C_s + \frac{1}{\left\{ 2C^s D^s \left[ \left( \frac{\alpha D^s}{2C^s} \pm \frac{1}{2C^s} \left( \sqrt{\frac{\alpha^2 (D^s)^2 + 4\sigma^s (D^s)^2}{+4\eta^s (C^s)^2 - 4\zeta^s C^s D^s}} \right) \right) + n\alpha \frac{D^s}{C^s} \right] \right\}^2}$$

$$\times \left\{ -\left[ \left( \frac{\alpha D^s}{2C^s} \pm \frac{1}{2C^s} \left( \sqrt{\frac{\alpha^2 (D^s)^2 + 4\sigma^s (D^s)^2}{+4\eta^s (C^s)^2 - 4\zeta^s C^s D^s}} \right) \right) + n\alpha \frac{D^s}{C^s} \right]^2 (C^s)^2 - \sigma^s (D^s)^2 + \eta^s (C^s)^2 \right\}^2 - \sigma^s = 0. \quad (50)$$

For this case, the upper component of the wave function is

$$F_{n\kappa}^s(r) = N_{n\kappa} \left( -\frac{D^s}{C^s} e^{-\alpha r} \right)^{\sqrt{\gamma_3}} \left( 1 + \frac{D^s}{C^s} e^{-\alpha r} \right)^{\frac{1}{2} + \sqrt{\gamma_1 - \gamma_2 + \gamma_3 + \frac{1}{4}}} P_n^{(2\sqrt{\gamma_3}, 2\sqrt{\gamma_1 - \gamma_2 + \gamma_3 + \frac{1}{4}})} (1 + \frac{2D^s}{C^s} e^{-\alpha r}), \qquad (51)$$

where

$$\gamma_1 = \frac{MV_0 B^2}{\alpha^2 D^2} + \frac{V_0 E_{n\kappa}^s B^{s2}}{\alpha^2 D^{s2}} - \frac{V_0 C_s B^{s2}}{\alpha^2 D^{s2}} - \frac{MV_1 C^s}{\alpha D^s} - \frac{E_{n\kappa}^s V_1 C^s}{\alpha D^s} + \frac{C_s V_1 C^s}{\alpha D^s} - \frac{M^2}{\alpha^2} + \frac{MC_s}{\alpha^2} + \frac{(E_{n\kappa}^s)^2}{\alpha^2} - \frac{E_{n\kappa}^s C_s}{\alpha^2},$$

$$\gamma_2 = \frac{2A^s B^s V_0 M}{\alpha^2 D^s C^s} + \frac{2A^s B^s V_0 E_{n\kappa}^s}{\alpha^2 D^s C^s} - \frac{2A^s B^s V_0 C_s}{\alpha^2 D^s C^s} - \frac{MV_1 C^s}{\alpha D^s} - \frac{E_{n\kappa}^s V_1 C^s}{\alpha D^s} + \frac{C_s V_1 C^s}{\alpha D^s} - \frac{2M^2}{\alpha^2} + \frac{2MC_s}{\alpha^2} + \frac{2(E_{n\kappa}^s)^2}{\alpha^2} - \frac{2E_{n\kappa}^s C_s}{\alpha^2},$$

$$\gamma_3 = (\kappa + H +)(\kappa + H) + \frac{MV_0 A^{s2}}{\alpha^2 C^{s2}} + \frac{V_0 E_{n\kappa}^s A^{s2}}{\alpha^2 C^{s2}} - \frac{V_0 C_s A^{s2}}{\alpha^2 C^{s2}} - \frac{M^2}{\alpha^2} + \frac{MC_s}{\alpha^2} + \frac{(E_{n\kappa}^s)^2}{\alpha^2} - \frac{E_{n\kappa}^s C_s}{\alpha^2}. \qquad (52)$$

and the other component can be simply found via

$$G_{n\kappa}^s(r) = \frac{1}{M + E_{n\kappa}^s - C_s}\left(\frac{d}{dr} + \frac{\kappa}{r} - U(r)\right)F_{n\kappa}^s(r), \qquad (53)$$

## 5. Solution of the Second Choice

### 5.1. Pseudospin Symmetry Limit for the Second Choice

The pseudospin symmetry occurs when $\dfrac{d\Sigma(r)}{dr} = 0$ or equivalently $\Sigma(r) = C_{ps} = Const.$. In this limit, we take $\Delta(r)$ as the MS-QY potential and pick up a Coulomb-like tensor potential:

$$\Delta(r) = V_0\left(\frac{A^{ps} + B^{ps}e^{-\alpha r}}{C^{ps} + D^{ps}e^{-\alpha r}}\right)^2 - V_1\left(1 - \frac{e^{-\alpha r}}{r}\right)^2, \qquad (54)$$

$$U(r) = -\frac{H}{r}, H = \frac{z_a z_b e^2}{4\pi\varepsilon_0}, r \geq R_\varepsilon \qquad (55)$$

Now, substitution of the proper approximation (see Fig. (1))

$$\frac{1}{r^2} = \lim_{\alpha \to 0}\left[\frac{C^{ps2}\alpha^2}{\left(C^{ps} + D^{ps}e^{-\alpha r}\right)^2}\right], \qquad (56)$$

in Eq. (22) yields

$$\{-\frac{d^2}{dr^2} + \frac{1}{(1 + \frac{D^{ps}}{C^{ps}}e^{-\alpha r})^2}\{[\frac{-MV_0 B^{ps2}}{C^{ps2}} + \frac{V_0 E_{n\kappa}^{ps} B^{ps2}}{C^{ps2}} - \frac{V_0 C_{ps} B^{ps2}}{C^{ps2}} + MV_1\alpha^2 - E_{n\kappa}^s V_1\alpha^2 + C_{ps}V_1\alpha^2 - \frac{2MV_1\alpha D^{ps}}{C^{ps}}$$

$$+ \frac{2E_{n\kappa}^{ps}V_1\alpha D^{ps}}{C^{ps}} - \frac{2C_{ps}V_1\alpha D^{ps}}{C^{ps}}]e^{-2\alpha r} + [-\frac{2A^{ps}B^{ps}V_0 M}{C^{ps2}} + \frac{2A^{ps}B^{ps}V_0 E_{n\kappa}^{ps}}{C^{ps2}} - \frac{2A^{ps}B^{ps}V_0 C_{ps}}{C^{ps2}} - 2MV_1\alpha + 2E_{n\kappa}^{ps}V_1\alpha$$

$$- 2C_{ps}V_1\alpha]e^{-\alpha r} + (\kappa + H)(\kappa + H - 1)\alpha^2 - \frac{MV_0 A^{ps2}}{C^{ps2}} + \frac{V_0 E_{n\kappa}^{ps} A^{ps2}}{C^{ps2}} - \frac{V_0 C_{ps} A^{ps2}}{C^{ps2}}\}G_{n\kappa}^{ps}(r) = \{-M^2 - MC_{ps} + (E_{n\kappa}^{ps})^2$$

$$- E_{n\kappa}^{ps} C_{ps} - MV_1 + E_{n\kappa}^s V_1 - C_{ps}V_1\}G_{n\kappa}^{ps}(r), \qquad (57)$$

where $\kappa = -\tilde{\ell}$ and $\kappa = \tilde{\ell} + 1$ for $\kappa < 0$ and $\kappa > 0$.

### 5.2. Solution of Pseudospin Symmetry Limit for the Second Choice

we obtain a Schrodinger-like equation of the form

$$-\frac{d^2 G_{n\kappa}^{ps}(r)}{dr^2} + V_{eff}(r)G_{n\kappa}^{ps}(r) = \tilde{E}_{n\kappa}^{ps}G_{n\kappa}^{ps}(r), \qquad (58)$$

with the effective potential being

$$V_{eff} = \frac{\mu^{ps}e^{-2\alpha r} + \lambda^{ps}e^{-\alpha r} + \chi^{ps}}{(1 + \frac{D}{C}e^{-\alpha r})^2}, \qquad (59)$$

Where

$$\mu^{ps} = \frac{-MV_0 B^{ps\,2}}{C^{ps\,2}} + \frac{V_0 E_{n\kappa}^{ps} B^{ps\,2}}{C^{ps\,2}} - \frac{V_0 C_{ps} B^{ps\,2}}{C^{ps\,2}} + MV_1\alpha^2 - E_{n\kappa}^s V_1\alpha^2 + C_{ps}V_1\alpha^2 - \frac{2MV_1\alpha D^{ps}}{C^{ps}} + \frac{2E_{n\kappa}^{ps}V_1\alpha D^{ps}}{C^{ps}} - C^{ps},$$

$$\lambda^{ps} = -\frac{2A^{ps}B^{ps}V_0 M}{C^{ps\,2}} + \frac{2A^{ps}B^{ps}V_0 E_{n\kappa}^{ps}}{C^{ps\,2}} - \frac{2A^{ps}B^{ps}V_0 C_{ps}}{C^{ps\,2}} - 2MV_1\alpha + 2E_{n\kappa}^{ps}V_1\alpha - 2C_{ps}V_1\alpha,$$

$$\chi^{ps} = (\kappa+H)(\kappa+H-1)\alpha^2 - \frac{MV_0 A^{ps\,2}}{C^{ps\,2}} + \frac{V_0 E_{n\kappa}^{ps} A^{ps\,2}}{C^{ps\,2}} - \frac{V_0 C_{ps} A^{ps\,2}}{C^{ps\,2}}. \qquad \textcolor{red}{(60)}$$

The corresponding effective energy is given by

$$\tilde{E}_{n\kappa}^{ps} = -M^2 - MC_{ps} + (E_{n\kappa}^{ps})^2 - E_{n\kappa}^{ps}C_{ps} - MV_1 + E_{n\kappa}^s V_1 - C_{ps}V_1, \qquad (61)$$

Thus, we are dealing with the Riccati equation

$$\phi(r) = \frac{q^{ps} e^{-\alpha r}}{\left(1 + \dfrac{D^{ps}}{C^{ps}} e^{-\alpha r}\right)} + w^{ps}. \qquad (62)$$

Substituting equation (62) into (32) yields

$$\tilde{E}_{0\kappa}^{ps} = \chi^{ps} - (w^{ps})^2, \qquad\qquad\qquad \textcolor{red}{(63-a)}$$

$$q^{ps} = \frac{\alpha D^{ps}}{2C^{ps}} - \frac{1}{2C^{ps}}\sqrt{\alpha^2(D^{ps})^2 + 4\chi^{ps}(D^{ps})^2 + 4\mu^{ps}(C^{ps})^2 - 4\lambda^{ps}C^{ps}D^{ps}}, \qquad \textcolor{red}{(63-b)}$$

$$w^{ps} = \frac{-(q^{ps})^2(C^{ps})^2 - \chi^{ps}(D^{ps})^2 + \mu^{ps}(C^{ps})^2}{2C^{ps}D^{ps}q^{ps}}. \qquad \textcolor{red}{(63-c)}$$

The corresponding supersymmetric partner potentials can be as

$$V_{eff_+}(r) = \phi^2 + \frac{d\phi}{dr}$$

$$= \frac{-q^{ps}\frac{C^{ps}}{D^{ps}}[q^{ps} + \alpha\frac{D^{ps}}{C^{ps}}]e^{-\alpha r}}{(1 + \frac{D^{ps}}{C^{ps}}e^{-\alpha r})^2} + \frac{[\frac{-\chi^{ps}(D^{ps})^2 + \mu^{ps}(C^{ps})^2}{C^{ps}D^{ps}}]e^{-\alpha r}}{(1 + \frac{D^{ps}}{C^{ps}}e^{-\alpha r})}$$

$$+ (\frac{-(q^{ps})^2(C^{ps})^2 - \chi^{ps}(D^{ps})^2 + \mu^{ps}(C^{ps})^2}{2C^{ps}D^{ps}q^{ps}})^2, \qquad (64-a)$$

$$V_{eff_-}(r) = \phi^2 - \frac{d\phi}{dr}$$

$$\frac{-q^{ps}\frac{C^{ps}}{D^{ps}}[q^{ps} - \alpha\frac{D^{ps}}{C^{ps}}]e^{-\alpha r}}{(1 + \frac{D^{ps}}{C^{ps}}e^{-\alpha r})^2} + \frac{[\frac{-\chi^{ps}(D^{ps})^2 + \mu^{ps}(C^{ps})^2}{C^{ps}D^{ps}}]e^{-\alpha r}}{(1 + \frac{D^{ps}}{C^{ps}}e^{-\alpha r})}$$

$$+ (\frac{-(q^{ps})^2(C^{ps})^2 - \chi^{ps}(D^{ps})^2 + \mu^{ps}(C^{ps})^2}{2C^{ps}D^{ps}q^{ps}})^2, \qquad (64-b)$$

where $a_0 = q^{ps}$ and $a_i$ is a function of $a_0$, i.e., $a_1 = f(a_0) = a_0 + \alpha\frac{D^{ps}}{C^{ps}}$. Consequently,

$a_n = f(a_0) = a_0 + n\alpha\frac{D^{ps}}{C^{ps}}$. We see that the shape invariance holds via a mapping of the form

$q^{ps} \to q^{ps} + \alpha\frac{D^{ps}}{C^{ps}}$. From Eq. (24), we have

$$R(a_1) = (\frac{-a_0^2(C^{ps})^2 - \chi^{ps}(D^{ps})^2 + \mu^{ps}(C^{ps})^2}{2C^{ps}D^{ps}a_0})^2 - (\frac{-a_1^2(C^{ps})^2 - \chi^{ps}(D^{ps})^2 + \mu^{ps}(C^{ps})^2}{2C^{ps}D^{ps}a_1})^2,$$

$$R(a_2) = (\frac{-a_1^2(C^{ps})^2 - \chi^{ps}(D^{ps})^2 + \mu^{ps}(C^{ps})^2}{2C^{ps}D^{ps}a_1})^2 - (\frac{-a_2^2(C^{ps})^2 - \chi^{ps}(D^{ps})^2 + \mu^{ps}(C^{ps})^2}{2C^{ps}D^{ps}a_2})^2,$$

$$R(a_3) = (\frac{-a_2^2(C^{ps})^2 - \chi^{ps}(D^{ps})^2 + \mu^{ps}(C^{ps})^2}{2C^{ps}D^{ps}a_2})^2 - (\frac{-a_3^2(C^{ps})^2 - \chi^{ps}(D^{ps})^2 + \mu^{ps}(C^{ps})^2}{2C^{ps}D^{ps}a_3})^2,$$

.

.

$$R(a_n) = (\frac{-a_{n-1}^2(C^{ps})^2 - \chi^{ps}(D^{ps})^2 + \mu^{ps}(C^{ps})^2}{2C^{ps}D^{ps}a_{n-1}})^2 - (\frac{-a_n^2(C^{ps})^2 - \chi^{ps}(D^{ps})^2 + \mu^{ps}(C^{ps})^2}{2C^{ps}D^{ps}a_n})^2, \qquad (65)$$

$$\tilde{E}_{0\kappa}^- = 0 \qquad (66)$$

Therefore, the from Eq. (65) the eigenvalues can be found as

$$\tilde{E}_{n\kappa}^{ps-} = \sum_{k=1}^{n} R(a_\kappa) = (\frac{-a_0^{\ 2}(C^{ps})^2 - \chi^{ps}(D^{ps})^2 + \mu^{ps}(C^{ps})^2}{2C^{ps}D^{ps}a_0})^2$$

$$-(\frac{-a_n^{\ 2}(C^{ps})^2 - \chi^{ps}(D^{ps})^2 + \mu^{ps}(C^{ps})^2}{2C^{ps}D^{ps}a_n})^2, \qquad (67-a)$$

$$\tilde{E}_{n\kappa}^{ps} = \tilde{E}_{n\kappa}^{ps-} + \tilde{E}_{0\kappa}^{ps} = -(\frac{-a_n^{\ 2}(C^{ps})^2 - \chi^{ps}(D^{ps})^2 + \mu^{ps}(C^{ps})^2}{2C^{ps}D^{ps}a_n})^2 + \chi^{ps}, \qquad (67-b)$$

By considering equations (59)–(62-c), the equation (67-b) becomes

$$-M^2 - MC_{ps} + (E_{n\kappa}^{ps})^2 - E_{n\kappa}^{ps}C_{ps} - MV_1 + E_{n\kappa}^s V_1 - C_{ps}V + \frac{1}{\left\{ 2C^{ps}D^{ps}\left[ \left( \frac{\alpha D^{ps}}{2C^{ps}} \pm \frac{1}{2C^{ps}}\left( \sqrt{\begin{matrix}\alpha^2(D^{ps})^2 + 4\chi^{ps}(D^{ps})^2 \\ +4\mu^{ps}(C^{ps})^2 - 4\lambda^{ps}C^{ps}D^{ps}\end{matrix}} \right) \right) + n\alpha\frac{D^{ps}}{C^{ps}} \right] \right\}^2}$$

$$\times \left\{ -\left[ \left( \frac{\alpha D^{ps}}{2C^{ps}} \pm \frac{1}{2C^{ps}}\left( \sqrt{\begin{matrix}\alpha^2(D^{ps})^2 + 4\chi^{ps}(D^{ps})^2 \\ +4\mu^{ps}(C^{ps})^2 - 4\lambda^{ps}C^{ps}D^{ps}\end{matrix}} \right) \right) + n\alpha\frac{D^{ps}}{C^{ps}} \right]^2 (C^{ps})^2 - \chi^{ps}(D^{ps})^2 + \mu^{ps}(C^{ps})^2 \right\}^2 - \chi^{ps} = 0. \quad (68)$$

In what follows, we find the lower component of the wave function as,

$$G_{n\kappa}^{ps}(r) = N_{n\kappa}\left( -\frac{D^{ps}}{C^{ps}}e^{-\alpha r} \right)^{\sqrt{\beta_3}}\left( 1 + \frac{D^{ps}}{C^{ps}}e^{-\alpha r} \right)^{\frac{1}{2} + \sqrt{\beta_1 - \beta_2 + \beta_3 + \frac{1}{4}}} P_n^{(2\sqrt{\beta_3}, 2\sqrt{\beta_1 - \beta_2 + \beta_3 + \frac{1}{4}})}(1 + \frac{2D^{ps}}{C^{ps}}e^{-\alpha r}), \qquad (69)$$

where

$$\beta_1 = -\frac{MV_0B^{ps2}}{\alpha^2 D^{ps2}} + \frac{V_0E_{n\kappa}^{ps}B^{ps2}}{\alpha^2 D^{ps2}} - \frac{V_0C_{ps}B^{ps2}}{\alpha^2 D^{ps2}} + \frac{MV_1C^{ps2}}{D^{ps2}} - \frac{E_{n\kappa}^{ps}V_1C^{ps2}}{D^{ps2}} + \frac{C_{ps}V_1C^{ps2}}{D^{ps2}} - \frac{2MV_1C^{ps}}{\alpha D^{ps}} + \frac{2E_{n\kappa}^{ps}V_1C^{ps}}{\alpha D^{ps}}$$

$$-\frac{2C_{ps}V_1C^{ps}}{\alpha D^{ps}} + \frac{M^2}{\alpha^2} + \frac{MC_{ps}}{\alpha^2} - \frac{(E_{n\kappa}^{ps})^2}{\alpha^2} + \frac{E_{n\kappa}^{ps}C_{ps}}{\alpha^2} + \frac{MV_1}{\alpha^2} - \frac{E_{n\kappa}^{ps}V_1}{\alpha^2} + \frac{C_{ps}V_1}{\alpha^2},$$

$$\beta_2 = -\frac{2A^{ps}B^{ps}V_0M}{\alpha^2 D^{ps}C^{ps}} + \frac{2AB^{ps}V_0E_{n\kappa}^{ps}}{\alpha^2 D^{ps}C^{ps}} - \frac{2AB^{ps}V_0C_{ps}}{\alpha^2 D^{ps}C^{ps}} - \frac{2MV_1C^{ps}}{\alpha D^{ps}} + \frac{2E_{n\kappa}^{ps}V_1C^{ps}}{\alpha D^{ps}} - \frac{2C_{ps}V_1C^{ps}}{\alpha D^{ps}} + \frac{2M^2}{\alpha^2} + \frac{2MC_{ps}}{\alpha^2}$$

$$-\frac{2(E_{n\kappa}^{ps})^2}{\alpha^2} + \frac{2E_{n\kappa}^{ps}C_{ps}}{\alpha^2} + \frac{2MV_1}{\alpha^2} - \frac{2E_{n\kappa}^{ps}V_1}{\alpha^2} + \frac{2C_{ps}V_1}{\alpha^2},$$

$$\beta_3 = (\kappa + H)(\kappa + H - 1) - \frac{MV_0A^{ps2}}{\alpha^2 C^{ps2}} + \frac{V_0E_{n\kappa}^{ps}A^{ps2}}{\alpha^2 C^{ps2}} - \frac{V_0C_{ps}A^{ps2}}{\alpha^2 C^{ps2}} + \frac{M^2}{\alpha^2} + \frac{MC_{ps}}{\alpha^2} - \frac{(E_{n\kappa}^{ps})^2}{\alpha^2} + \frac{E_{n\kappa}^{ps}C_{ps}}{\alpha^2} + \frac{MV_1}{\alpha^2}$$

$$-\frac{E_{n\kappa}^{ps}V_1}{\alpha^2} + \frac{C_{ps}V_1}{\alpha^2}. \qquad (70)$$

where $N_{n\kappa}$ is the normalization constant. The upper spinor component of the Dirac equation can be calculated as ,

$$F_{n\kappa}^{ps}(r) = \frac{1}{M - E_{n\kappa}^{ps} + C_{ps}}\left( \frac{d}{dr} - \frac{\kappa}{r} + U(r) \right)G_{n\kappa}^{ps}(r), \quad (71)$$

where $E_{n\kappa} \neq M + C_{ps}$ and when $C_{ps} = 0$ (exact pseudospin symmetry) is obtained which means that only negative energy solutions are possible.

## 5.3 Spin Symmetry Limit for the Second Choice

Now, we investigate the analytical solution of the Dirac equation in the spin symmetry limit

$$\Sigma(r) = V_0 \left( \frac{A^s + B^s e^{-\alpha r}}{C^s + D^s e^{-\alpha r}} \right)^2 - V_1 \left( \frac{1 - e^{-\alpha r}}{r} \right)^2, \qquad (72)$$

Substitution of the latter in Eq. (21) gives

$$\{ -\frac{d^2}{dr^2} + \frac{1}{(1 + \frac{D^s}{C^s} e^{-\alpha r})^2} \{ [\frac{V_0 E_{n\kappa}^s B^{s2}}{C^{s2}} - E_{n\kappa}^s V_1 \alpha^2 + \frac{MV_0 B^{s2}}{C^{s2}} - MV_1 \alpha^2 - \frac{V_0 C_s B^{s2}}{C^{s2}} + C_s V_1 \alpha^2 + \frac{2MV_1 \alpha D^s}{C^s}$$

$$+ \frac{2E_{n\kappa}^s V_1 \alpha D^s}{C^s} - \frac{2C_s V_1 \alpha D^s}{C^s} ] e^{-2\alpha r} + [\frac{2A^s B^s V_0 M}{C^{s2}} + \frac{2A^s B^s V_0 E_{n\kappa}^s}{C^{s2}} - \frac{2A^s B^s V_0 C_s}{C^{s2}} + 2MV_1 \alpha + 2E_{n\kappa}^s V_1 \alpha - 2C_s V_1 \alpha ] e^{-\alpha r}$$

$$+ (\kappa + H + 1)(\kappa + H) \alpha^2 + \frac{MV_0 A^{s2}}{C^{s2}} + \frac{V_0 E_{n\kappa}^s A^{s2}}{C^{s2}} - \frac{V_0 C_s A^{s2}}{C^{s2}} \} F_{n\kappa}^s(r) = \{ (E_{n\kappa}^s)^2 + E_{n\kappa}^s V_1 - M^2 + MV_1 - E_{n\kappa}^s C_s$$

$$+ MC_s - C_s V_1 \} F_{n\kappa}^s(r), \qquad (73)$$

where $\kappa = \ell$ and $\kappa = -\ell - 1$ for $\kappa < 0$ and $\kappa > 0$.

## 5.4. Solution of the Spin Symmetry Limit for the second choice

In this case,

$$-\frac{d^2 F_{n\kappa}^s(r)}{dr^2} + V_{eff}(r) F_{n\kappa}^s(r) = \tilde{E}_{n\kappa}^s F_{n\kappa}^s(r), \qquad (74)$$

with

$$V_{eff} = \frac{\mu^s e^{-2\alpha r} + \lambda^s e^{-\alpha r} + \chi^s}{(1 + \frac{D^s}{C^s} e^{-\alpha r})^2}, \qquad (75)$$

where

$$\mu^s = \frac{V_0 E_{n\kappa}^s B^{s2}}{C^{s2}} - E_{n\kappa}^s V_1 \alpha^2 + \frac{MV_0 B^{s2}}{C^{s2}} - MV_1 \alpha^2 - \frac{V_0 C_s B^{s2}}{C^{s2}} + C_s V_1 \alpha^2 + \frac{2MV_1 \alpha D^s}{C^s} + \frac{2E_{n\kappa}^s V_1 \alpha D^s}{C^s} - \frac{2C_s V_1 \alpha D^s}{C^s},$$

$$\lambda^s = \frac{2A^s B^s V_0 M}{C^{s2}} + \frac{2A^s B^s V_0 E_{n\kappa}^s}{C^{s2}} - \frac{2A^s B^s V_0 C_s}{C^{s2}} + 2MV_1 \alpha + 2E_{n\kappa}^s V_1 \alpha - 2C_s V_1 \alpha,$$

$$\chi^s = (\kappa + H + 1)(\kappa + H) \alpha^2 + \frac{MV_0 A^{s2}}{C^{s2}} + \frac{V_0 E_{n\kappa}^s A^{s2}}{C^{s2}} - \frac{V_0 C_s A^{s2}}{C^{s2}}. \qquad (76)$$

and

$$\tilde{E}_{n\kappa}^s = (E_{n\kappa}^s)^2 + E_{n\kappa}^s V_1 - M^2 + MV_1 - E_{n\kappa}^s C_s + MC_s - C_s V_1, \qquad (77)$$

Energy eigenvalues equation for the MS-QY potential potential in the presence of tensor interaction in view of the spin symmetry limit is obtained as follows

$$(E_{n\kappa}^s)^2 + E_{n\kappa}^s V_1 - M^2 + MV_1 - E_{n\kappa}^s C_s + MC_s - C_s V_1 + \frac{1}{\left\{ 2C^s D^s \left[ \left( \frac{\alpha D^s}{2C^s} \pm \frac{1}{2C^s} \left( \sqrt{\alpha^2 (D^s)^2 + 4\chi^s (D^s)^2 \atop + 4\mu^s (C^s)^2 - 4\lambda^s C^s D^s} \right) \right) + n\alpha \frac{D^s}{C^s} \right] \right\}^2}$$

$$\times \left\{ -\left[ \left( \frac{\alpha D^s}{2C^s} \pm \frac{1}{2C^s} \left( \sqrt{\alpha^2 (D^s)^2 + 4\chi^s (D^s)^2 \atop + 4\mu^s (C^s)^2 - 4\lambda^s C^s D^s} \right) \right) + n\alpha \frac{D^s}{C^s} \right]^2 (C^s)^2 - \chi^{ps} (D^s)^2 + \mu^s (C^s)^2 \right\}^2 - \chi^s = 0. \quad (78)$$

$$F_{n\kappa}^s(r) = N_{n\kappa} \left( -\frac{D^s}{C^s} e^{-\alpha r} \right)^{\sqrt{\nu_3}} \left( 1 + \frac{D^s}{C^s} e^{-\alpha r} \right)^{\frac{1}{2} + \sqrt{\nu_1 - \nu_2 + \nu_3 + \frac{1}{4}}} P_n^{(2\sqrt{\nu_3}, 2\sqrt{\nu_1 - \nu_2 + \nu_3 + \frac{1}{4}})} (1 + \frac{2D^s}{C^s} e^{-\alpha r}), \quad (79)$$

where

$$\nu_1 = \frac{V_0 E_{n\kappa}^s B^{s2}}{\alpha^2 D^{s2}} - \frac{E_{n\kappa}^s V_1 C^{s2}}{D^{s2}} + \frac{MV_0 B^{s2}}{\alpha^2 D^{s2}} - \frac{MV_1 C^{s2}}{D^{s2}} - \frac{V_0 C_s B^{s2}}{\alpha^2 D^{s2}} + \frac{C_s V_1 C^{s2}}{D^{s2}} + \frac{2MV_1 C^s}{\alpha D^s} + \frac{2E_{n\kappa}^s V_1 C^s}{\alpha D^s} - \frac{2C_s V_1 C^s}{\alpha D^s} - \frac{(E_{n\kappa}^s)^2}{\alpha^2}$$

$$- \frac{E_{n\kappa}^s V_1}{\alpha^2} + \frac{M^2}{\alpha^2} - \frac{MV_1}{\alpha^2} + \frac{E_{n\kappa}^s C_s}{\alpha^2} - \frac{MC_s}{\alpha^2} + \frac{C_s V_1}{\alpha^2},$$

$$\nu_2 = \frac{2A^s B^s V_0 M}{\alpha^2 D^s C^s} + \frac{2A^s B^s V_0 E_{n\kappa}^s}{\alpha^2 D^s C^s} - \frac{2A^s B^s V_0 C_s}{\alpha^2 D^s C^s} + \frac{2MV_1 C^s}{\alpha D^s} + \frac{2E_{n\kappa}^s V_1 C^s}{\alpha D^s} - \frac{2C_s V_1 C^s}{\alpha D^s} - \frac{2(E_{n\kappa}^s)^2}{\alpha^2} - \frac{2E_{n\kappa}^s V_1}{\alpha^2} + \frac{2M^2}{\alpha^2} - \frac{2MV_1}{\alpha^2}$$

$$+ \frac{2E_{n\kappa}^s C_s}{\alpha^2} - \frac{2MC_s}{\alpha^2} + \frac{2C_s V_1}{\alpha^2},$$

$$\nu_3 = (\kappa + H + 1)(\kappa + H) + \frac{MV_0 A^{s2}}{\alpha^2 C^{s2}} + \frac{V_0 E_{n\kappa}^s A^{s2}}{\alpha^2 C^{s2}} - \frac{V_0 C_s A^{s2}}{\alpha^2 C^{s2}} - \frac{(E_{n\kappa}^s)^2}{\alpha^2} - \frac{E_{n\kappa}^{ps} V_1}{\alpha^2} + \frac{M^2}{\alpha^2} - \frac{MV_1}{\alpha^2} + \frac{E_{n\kappa}^s C_s}{\alpha^2} - \frac{MC_s}{\alpha^2} + \frac{C_s V_1}{\alpha^2}. \quad (80)$$

and the other component can be simply found via

$$G_{n\kappa}^s(r) = \frac{1}{M + E_{n\kappa}^s - C_s} (\frac{d}{dr} + \frac{\kappa}{r} - U(r)) F_{n\kappa}^s(r), \qquad (81)$$

## 6. Few special cases

In this section, we will consider some special cases of interest of the MS-Yukawa potential and MS-QY potential as follows. Maghsoodi et al. [31] have obtained the approximate solutions of the Dirac equation in the presence of Deng-Fan potential plus a tensor interaction term using the SUSSQM formalism .If we set $V_1 = 0$, the potential reduces into the Mobius square potential. The generalized form of the Mobius square can be written as,

$$V(r) = \frac{V_0}{C^2} \left( \frac{A^2 + 2ABe^{-\alpha r} + B^2 e^{-2\alpha r}}{\left(1 + \dfrac{D}{C} e^{-\alpha r}\right)^2} \right) . \qquad (82)$$

Comparing Eq.(46) with Deng-Fan potential [31],

$$V(r) = D_e \left( \frac{1 - 2(1+b)e^{-\alpha r} + (1+b)^2 e^{-2\alpha r}}{\left(1 - e^{-\alpha r}\right)^2} \right) , \qquad (83)$$

for $C = 1, D' = -1$, we have $V_0 A^2 = D_e, V_0 AB = -D_e(1+b), V_0 B^2 = D_e(1+b)^2$, where $b = e^{\alpha r_d} - 1$ and $r_d$ is the equilibrium inter-nuclear distance. This result is consistent with that of Maghsoodi et al. [31] using SUSYQM. In [34], approximate solutions of the Dirac equation in the presence of Yukawa potential plus a tensor interaction term is reported. If we set $V_0 = 0$, for the first choice, we obtain the Yukawa potential [34,35]. This result is consistent [34]. Finally, when $C_{ps} = 0$ and $\alpha \to 0$, Eq.(43) yields the energy formula for the Coulomb-like potential as [36]. In tables (1)-(4) we have portrayed the energy values for various $H$ for pseudospin and spin symmetry limits. Figs. (2)-(5) show energy vs. $\alpha$, $V_0$, $V_1$ and $H$ for pseudospin and spin symmetry limits for the first and second choices, respectively. Figs. (6)-(8) show energy vs. $\alpha$, $V_0$ and $H$ for pseudospin and spin symmetry limits. In Fig. (9) and Fig. (10), the wave functions are plotted for spin and pseudospin symmetry limits for the first and second choices of the potential with and without a tensor Interaction, respectively.

## 7 Conclusions

In this paper, we have obtained the approximate solutions of the Dirac equation for the MS-Y potential and the MS-QY potential including the tensor Coulomb interaction within the framework of pseudospin and spin symmetry limits using the SUSYQM which stands a strong tool of mathematical physics. We have obtained the energy eigenvalues and corresponding lower and upper wave functions in terms of the Jacobi polynomials. Moreover, the results obtained in this work have been compared with the previous work of other authors previously released.

**Table (1).** Energies in the Pseudospin Symmetry Limit for the first choice

$\alpha$=0.01 ,  M=5fm$^{-1}$,$A^{ps}$= 1, $B^{ps}$ = $-2$, $C^{ps}$ = 1, $D^{ps}$ = $-1$,C$_{ps}$=0, $V_0$ = $-0.2$, $V_1$ = 0.1.

| $\tilde{\ell}$ | $n, \kappa < 0$ | $(\ell, j)$ | $E_{n\kappa}^{ps}(fm^{-1})$ $(H=0)$ | $E_{n\kappa}^{ps}(fm^{-1})$ $(H=0.5)$ | $n-1, \kappa > 0$ | $(\ell+2, j+1)$ | $E_{n\kappa}^{ps}(fm^{-1})$ $(H=0)$ | $E_{n\kappa}^{ps}(fm^{-1})$ $(H=0.5)$ |
|---|---|---|---|---|---|---|---|---|
| 1 | 1,-1 | $1S_{\frac{1}{2}}$ | -5.009375979 | -5.009327474 | 0,2 | $0d_{\frac{3}{2}}$ | -5.009375979 | -5.009443876 |
| 2 | 1,-2 | $1P_{\frac{3}{2}}$ | -5.009531153 | -5.009443876 | 0,3 | $0f_{\frac{5}{2}}$ | -5.009531153 | -5.009637797 |
| 3 | 1,-3 | $1d_{\frac{5}{2}}$ | -5.00976379 | -5.009637797 | 0,4 | $0g_{\frac{7}{2}}$ | -5.00976379 | -5.009909113 |
| 4 | 1,-4 | $1f_{\frac{7}{2}}$ | -5.010073741 | -5.009909113 | 0,5 | $0h_{\frac{9}{2}}$ | -5.010073741 | -5.01025765 |
| 1 | 2,-1 | $2S_{\frac{1}{2}}$ | -5.014732692 | -5.01468506 | 1,2 | $1d_{\frac{3}{2}}$ | -5.014732692 | -5.014799366 |
| 2 | 2,-2 | $2P_{\frac{3}{2}}$ | -5.014885072 | -5.014799366 | 1,3 | $1f_{\frac{5}{2}}$ | -5.014885072 | -5.014989796 |

| ℓ | $n,\kappa<0$ | $(\ell,j)$ | $E_{n\kappa}^{s}(fm^{-1})$ $(H=0)$ | $E_{n\kappa}^{s}(fm^{-1})$ $(H=0.5)$ | $n,\kappa>0$ | $(\ell,j)$ | $E_{n\kappa}^{s}(fm^{-1})$ $(H=0)$ | $E_{n\kappa}^{s}(fm^{-1})$ $(H=0.5)$ |
|---|---|---|---|---|---|---|---|---|
| 3 | 2,-3 | $2d_{\frac{5}{2}}$ | -5.015113521 | -5.014989796 | 1,4 | $1g_{\frac{7}{2}}$ | -5.015113521 | -5.015256229 |
| 4 | 2,-4 | $2f_{\frac{7}{2}}$ | -5.015417895 | -5.015256229 | 1,5 | $1h_{\frac{9}{2}}$ | -5.015417895 | -5.015598495 |
| 1 | 3,-1 | $3S_{\frac{1}{2}}$ | -5.019956641 | -5.019909859 | 2,2 | $2d_{\frac{3}{2}}$ | -5.019956641 | -5.020022126 |
| 2 | 3,-2 | $3P_{\frac{3}{2}}$ | -5.020106303 | -5.020022126 | 2,3 | $2f_{\frac{5}{2}}$ | -5.020106303 | -5.020209158 |
| 3 | 3,-3 | $3d_{\frac{5}{2}}$ | -5.020330677 | -5.020209158 | 2,4 | $2g_{\frac{7}{2}}$ | -5.020330677 | -5.020470839 |
| 4 | 3,-4 | $3f_{\frac{7}{2}}$ | -5.020629623 | -5.020470839 | 2,5 | $2h_{\frac{9}{2}}$ | -5.020629623 | -5.020807004 |

**Table (2).** Energies
$\alpha$=0.01 , M=5fm$^{-1}$, $A^{s}=1$, $B^{s}=-2$, $C^{s}=1$, $D^{s}=-1$, C$_{s}=0$, $V_{0}=0.2$, $V_{1}=0.1$.

| ℓ | $n,\kappa<0$ | $(\ell,j)$ | $E_{n\kappa}^{s}(fm^{-1})$ $(H=0)$ | $E_{n\kappa}^{s}(fm^{-1})$ $(H=0.5)$ | $n,\kappa>0$ | $(\ell,j)$ | $E_{n\kappa}^{s}(fm^{-1})$ $(H=0)$ | $E_{n\kappa}^{s}(fm^{-1})$ $(H=0.5)$ |
|---|---|---|---|---|---|---|---|---|
| 1 | 0, -2 | $0P_{\frac{3}{2}}$ | 5.001904476 | 5.001854816 | 0,1 | $0P_{\frac{1}{2}}$ | 5.001904476 | 5.001973989 |
| 2 | 0, -3 | $0d_{\frac{5}{2}}$ | 5.002063344 | 5.001973989 | 0,2 | $0d_{\frac{3}{2}}$ | 5.002063344 | 5.002172527 |
| 3 | 0, -4 | $0f_{\frac{7}{2}}$ | 5.002301519 | 5.002172527 | 0,3 | $0f_{\frac{5}{2}}$ | 5.002301519 | 5.0024503 |
| 4 | 0,-5 | $0g_{\frac{9}{2}}$ | 5.002618847 | 5.0024503 | 0,4 | $0g_{\frac{7}{2}}$ | 5.002618847 | 5.002807131 |
| 1 | 1, -2 | $1P_{\frac{3}{2}}$ | 5.007439826 | 5.007391068 | 1,1 | $1P_{\frac{1}{2}}$ | 5.007439826 | 5.007508074 |
| 2 | 1, -3 | $1d_{\frac{5}{2}}$ | 5.007595804 | 5.007508074 | 1,2 | $1d_{\frac{3}{2}}$ | 5.007595804 | 5.007703001 |

| 3 | 1, -4 | $1f_{\frac{7}{2}}$ | 5.007829648 | 5.007703001 | 1,3 | $1f_{\frac{5}{2}}$ | 5.007829648 | 5.007975724 |
|---|---|---|---|---|---|---|---|---|
| 4 | 1, -5 | $1g_{\frac{9}{2}}$ | 5.008141207 | 5.007975724 | 1,4 | $1g_{\frac{7}{2}}$ | 5.008141207 | 5.008326069 |
| 1 | 2, -2 | $2P_{\frac{3}{2}}$ | 5.01283763 | 5.012789751 | 2,1 | $2P_{\frac{1}{2}}$ | 5.01283763 | 5.012904649 |
| 2 | 2, -3 | $2d_{\frac{5}{2}}$ | 5.012990798 | 5.012904649 | 2,2 | $2d_{\frac{3}{2}}$ | 5.012990798 | 5.013096063 |
| 3 | 2, -4 | $2f_{\frac{7}{2}}$ | 5.013220428 | 5.013096063 | 2,3 | $2f_{\frac{5}{2}}$ | 5.013220428 | 5.013363873 |
| 4 | 2, -5 | $2g_{\frac{9}{2}}$ | 5.013526376 | 5.013363873 | 2,4 | $2g_{\frac{7}{2}}$ | 5.013526376 | 5.01370791 |

**Table (3).** Energies in the Pseudospin Symmetry Limit for the second choice $\alpha = 0.01$ , M=5fm$^{-1}$, $A^{ps} = 1$, $B^{ps} = -2$, $C^{ps} = 1$, $D^{ps} = -1$, $C_{ps} = 0$, $V_0 = -0.2$, $V_1 = 0.1$.

| $\tilde{\ell}$ | $n, \kappa < 0$ | $(\ell, j)$ | $E_{n\kappa}^{ps}(fm^{-1})$ $(H = 0)$ | $E_{n\kappa}^{ps}(fm^{-1})$ $(H = 0.5)$ | $n-1, \kappa > 0$ | $(\ell+2, j+1)$ | $E_{n\kappa}^{ps}(fm^{-1})$ $(H = 0)$ | $E_{n\kappa}^{ps}(fm^{-1})$ $(H = 0.5)$ |
|---|---|---|---|---|---|---|---|---|
| 1 | 1,-1 | $1S_{\frac{1}{2}}$ | -5.106436115 | -5.106387711 | 0,2 | $0d_{\frac{3}{2}}$ | -5.106436115 | -5.10650387 |
| 2 | 1,-2 | $1P_{\frac{3}{2}}$ | -5.106590965 | -5.10650387 | 0,3 | $0f_{\frac{5}{2}}$ | -5.106590965 | -5.106697386 |
| 3 | 1,-3 | $1d_{\frac{5}{2}}$ | -5.106823116 | -5.106697386 | 0,4 | $0g_{\frac{7}{2}}$ | -5.106823116 | -5.106968137 |
| 4 | 1,-4 | $1f_{\frac{7}{2}}$ | -5.107132424 | -5.106968137 | 0,5 | $0h_{\frac{9}{2}}$ | -5.107132424 | -5.107315951 |
| 1 | 2,-1 | $2S_{\frac{1}{2}}$ | -5.11182877 | -5.111781234 | 1,2 | $1d_{\frac{3}{2}}$ | -5.11182877 | -5.111895309 |
| 2 | 2,-2 | $2P_{\frac{3}{2}}$ | -5.111980842 | -5.111895309 | 1,3 | $1f_{\frac{5}{2}}$ | -5.111980842 | -5.112085355 |
| 3 | 2,-3 | $2d_{\frac{5}{2}}$ | -5.112208832 | -5.112085355 | 1,4 | $1g_{\frac{7}{2}}$ | -5.112208832 | -5.112351252 |

| $\ell$ | $n,\kappa<0$ | $(\ell,j)$ | | | $n,\kappa>0$ | $(\ell,j)$ | | |
|---|---|---|---|---|---|---|---|---|
| 4 | 2,-4 | $2f_{\frac{7}{2}}$ | -5.112512595 | -5.112351252 | 1,5 | $1h_{\frac{9}{2}}$ | -5.112512595 | -5.112692834 |
| 1 | 3,-1 | $3S_{\frac{1}{2}}$ | -5.11708862 | -5.117041929 | 2,2 | $2d_{\frac{3}{2}}$ | -5.11708862 | -5.117153977 |
| 2 | 3,-2 | $3P_{\frac{3}{2}}$ | -5.11723799 | -5.117153977 | 2,3 | $2f_{\frac{5}{2}}$ | -5.11723799 | -5.117340646 |
| 3 | 3,-3 | $3d_{\frac{5}{2}}$ | -5.117461929 | -5.117340646 | 2,4 | $2g_{\frac{7}{2}}$ | -5.117461929 | -5.117601819 |
| 4 | 3,-4 | $3f_{\frac{7}{2}}$ | -5.117760296 | -5.117601819 | 2,5 | $2h_{\frac{9}{2}}$ | -5.117760296 | -5.117937335 |

**Table (4).** Energies
$\alpha$=0.01 , M=5fm$^{-1}$, $A^s=1$, $B^s=-2$, $C^s=1$, $D^s=-1$, $\mathrm{C}_s=0$, $V_0=0.2$, $V_1=0.1$.

| $\ell$ | $n,\kappa<0$ | $(\ell,j)$ | $E_{n\kappa}^{s}(fm^{-1})$ $(H=0)$ | $E_{n\kappa}^{s}(fm^{-1})$ $(H=0.5)$ | $n,\kappa>0$ | $(\ell,j)$ | $E_{n\kappa}^{s}(fm^{-1})$ $(H=0)$ | $E_{n\kappa}^{s}(fm^{-1})$ $(H=0.5)$ |
|---|---|---|---|---|---|---|---|---|
| 1 | 0, -2 | $0P_{\frac{3}{2}}$ | 4.904873061 | 4.904823288 | 0,1 | $0P_{\frac{1}{2}}$ | 4.904873061 | 4.904942731 |
| 2 | 0, -3 | $0d_{\frac{5}{2}}$ | 4.905032288 | 4.904942731 | 0,2 | $0d_{\frac{3}{2}}$ | 4.905032288 | 4.905141716 |
| 3 | 0, -4 | $0f_{\frac{7}{2}}$ | 4.905270998 | 4.905141716 | 0,3 | $0f_{\frac{5}{2}}$ | 4.905270998 | 4.905420113 |
| 4 | 0,-5 | $0g_{\frac{9}{2}}$ | 4.905589037 | 4.905420113 | 0,4 | $0g_{\frac{7}{2}}$ | 4.905589037 | 4.905777742 |
| 1 | 1, -2 | $1P_{\frac{3}{2}}$ | 4.910372527 | 4.910323664 | 1,1 | $1P_{\frac{1}{2}}$ | 4.910372527 | 4.910440925 |
| 2 | 1, -3 | $1d_{\frac{5}{2}}$ | 4.910528846 | 4.910440925 | 1,2 | $1d_{\frac{3}{2}}$ | 4.910528846 | 4.910636276 |
| 3 | 1, -4 | $1f_{\frac{7}{2}}$ | 4.910763198 | 4.910636276 | 1,3 | $1f_{\frac{5}{2}}$ | 4.910763198 | 4.910909592 |
| 4 | 1, -5 | $1g_{\frac{9}{2}}$ | 4.911075433 | 4.910909592 | 1,4 | $1g_{\frac{7}{2}}$ | 4.911075433 | 4.911260694 |
| 1 | 2, -2 | $2P_{\frac{3}{2}}$ | 4.915734463 | 4.915686483 | 2,1 | $2P_{\frac{1}{2}}$ | 4.915734463 | 4.915801623 |
| 2 | 2, -3 | $2d_{\frac{5}{2}}$ | 4.915887953 | 4.915801623 | 2,2 | $2d_{\frac{3}{2}}$ | 4.915887953 | 4.91599344 |
| 3 | 2, -4 | $2f_{\frac{7}{2}}$ | 4.916118066 | 4.91599344 | 2,3 | $2f_{\frac{5}{2}}$ | 4.916118066 | 4.916261812 |
| 4 | 2, -5 | $2g_{\frac{9}{2}}$ | 4.916424654 | 4.916261812 | 2,4 | $2g_{\frac{7}{2}}$ | 4.916424654 | 4.916606567 |

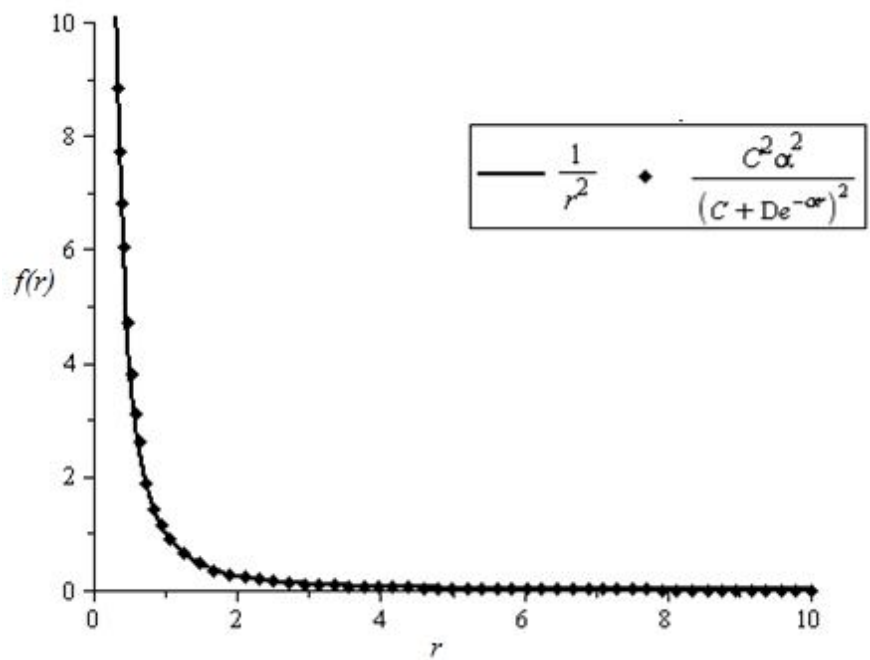

**Fig.(1).** The centrifugal term $(1/r^2)$ and its approximation for $\alpha = 0.01$.

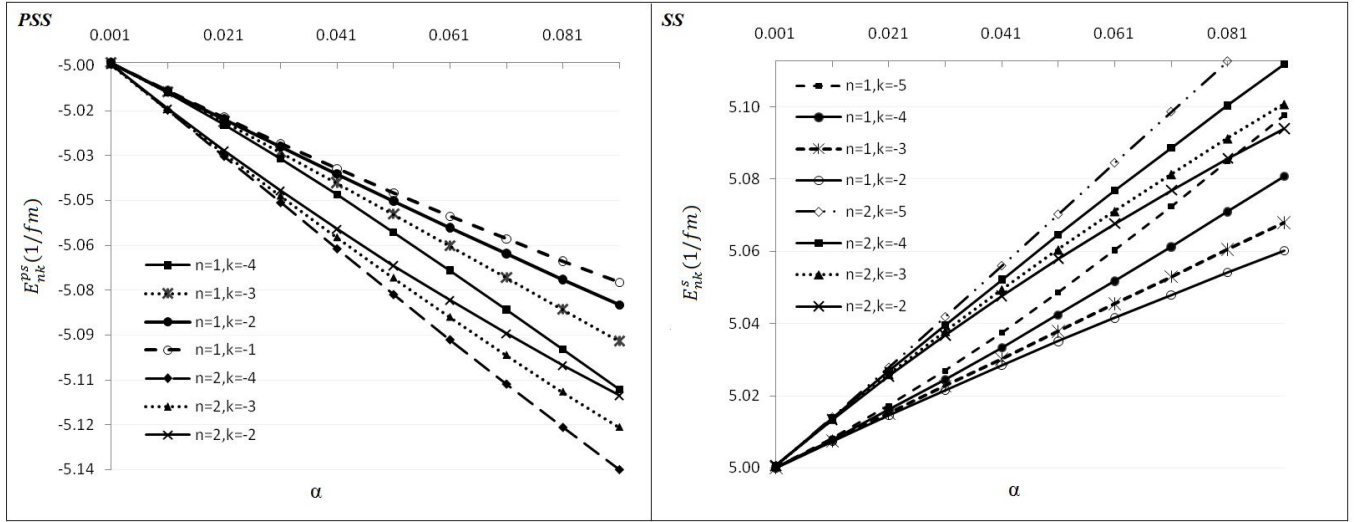

**Fig. (2).** *PSS*: Energy vs. $\alpha$ for Pseudospin Symmetry Limit for first choice
$H = 0.5$, M=5fm$^{-1}$, $A^{ps} = 1$, $B^{ps} = -2$, $C^{ps} = 1$, $D^{ps} = -1$, $C_{ps} = 0$, $V_0 = -0.2$, $V_1 = 0.1$.

   *SS:* Energy vs. $\alpha$ for Spin Symmetry Limit for first choice
$H = 0.5$, M=5fm$^{-1}$, $A^{s} = 1$, $B^{s} = -2$, $C^{s} = 1$, $D^{s} = -1$, $C_{s} = 0$, $V_0 = 0.2$, $V_1 = 0.1$.

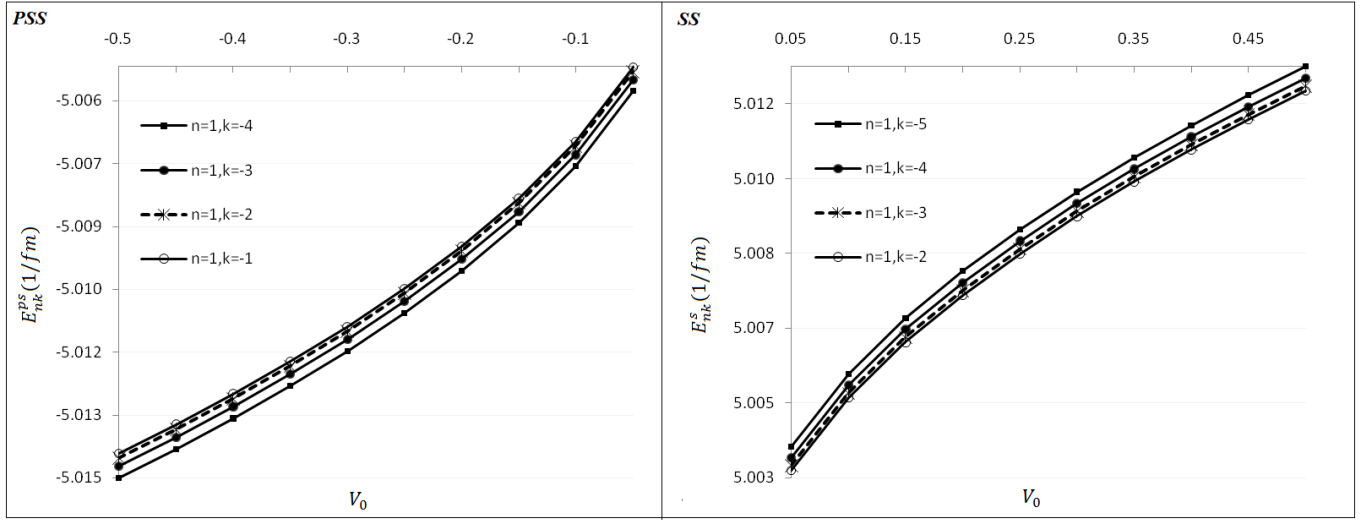

**Fig. (3).** *PSS*: Energy vs. $V_0$ for Pseudospin Symmetry Limit for first choice
$H = 0.5$, $\alpha$=0.01, M=5fm$^{-1}$, $A^{ps} = 1$, $B^{ps} = -2$, $C^{ps} = 1$, $D^{ps} = -1$, $C_{ps} = 0$, $V_1 = 0.1$.

   *SS:* Energy vs. $V_0$ for Spin Symmetry Limit for first choice
$H = 0.5$, $\alpha$=0.01, M=5fm$^{-1}$, $A^{s} = 1$, $B^{s} = -2$, $C^{s} = 1$, $D^{s} = -1$, $C_{s} = 0$, $V_1 = 0.1$.

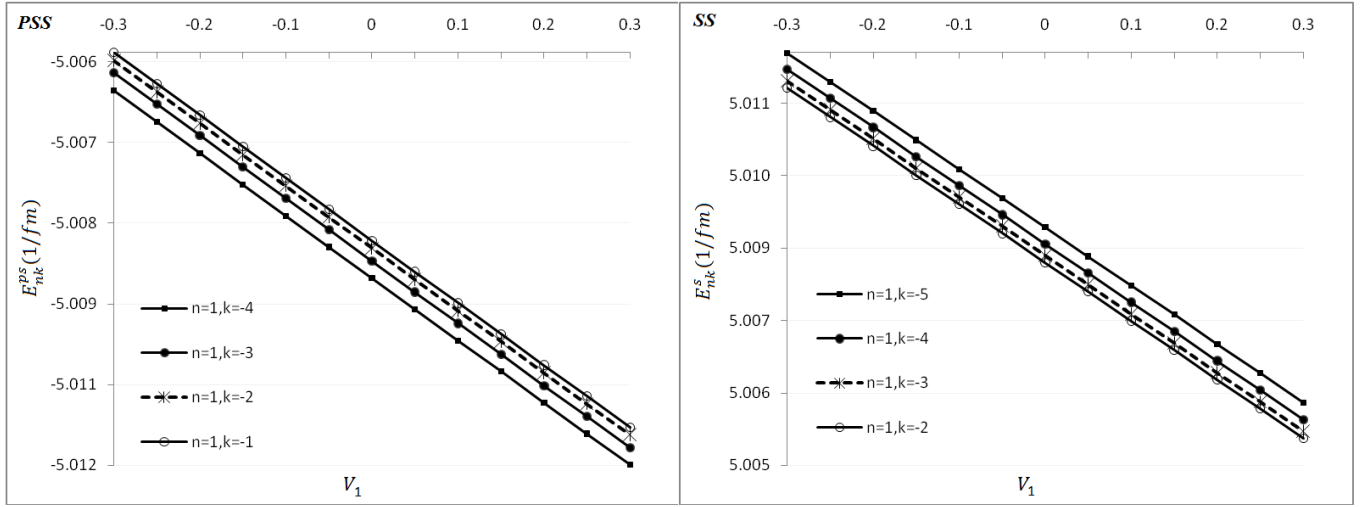

**Fig. (4).** ***PSS***: Energy vs. $V_1$ for Pseudospin Symmetry Limit for first choice

$H = 0.5, \alpha = 0.01$, M=5fm$^{-1}$, $A^{ps} = 1$, $B^{ps} = -2$, $C^{ps} = 1$, $D^{ps} = -1$, $C_{ps} = 0$, $V_0 = -0.2$.

***SS***: Energy vs. $V_1$ for Spin Symmetry Limit for first choice

$H = 0.5, \alpha = 0.01$, M=5fm$^{-1}$, $A^s = 1$, $B^s = -2$, $C^s = 1$, $D^s = -1$, $C_s = 0$, $V_0 = 0.2$.

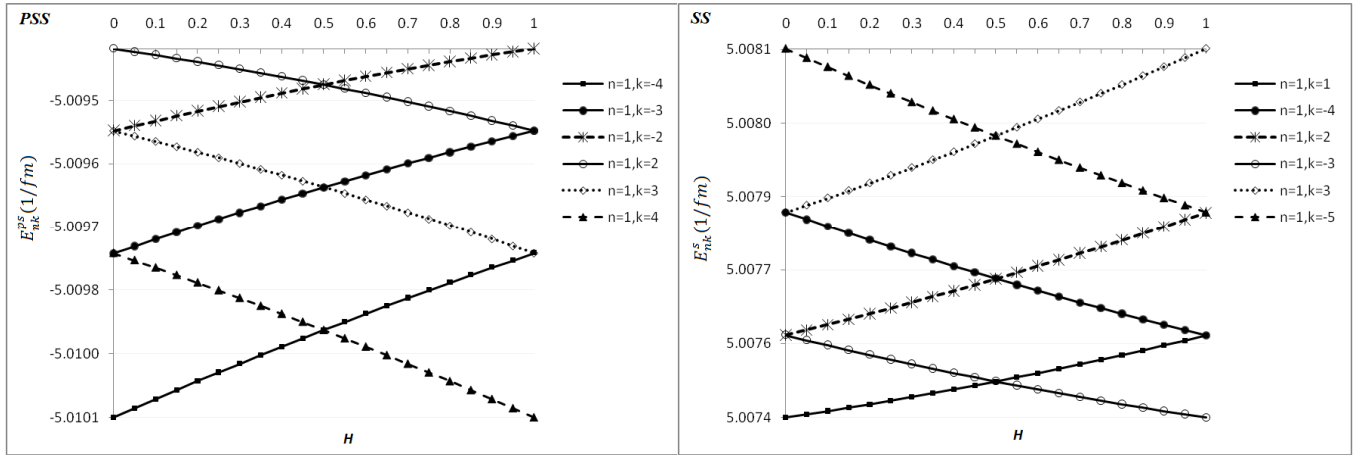

**Fig. (5).** ***PSS***: Energy vs. $H$ for Pseudospin Symmetry Limit for first choice

$\alpha = 0.01$, M=5fm$^{-1}$, $A^{ps} = 1$, $B^{ps} = -2$, $C^{ps} = 1$, $D^{ps} = -1$, $C_{ps} = 0$, $V_0 = -0.2$, $V_1 = 0.1$.

***SS***: Energy vs. $H$ for Spin Symmetry Limit for first choice

$\alpha = 0.01$, M=5fm$^{-1}$, $A^s = 1$, $B^s = -2$, $C^s = 1$, $D^s = -1$, $C_s = 0$, $V_0 = 0.2$, $V_1 = 0.1$.

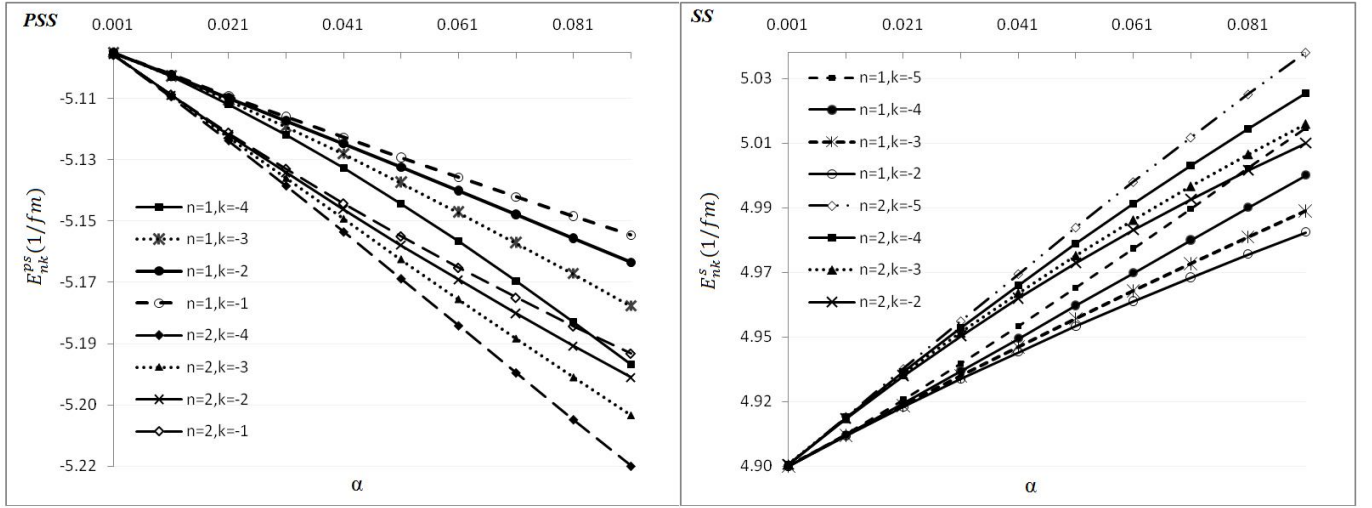

**Fig. (6).** *PSS*: Energy vs. $\alpha$ for Pseudospin Symmetry Limit for second choice
$H = 0.5$ , M=5fm$^{-1}$, $A^{ps} = 1$, $B^{ps} = -2$, $C^{ps} = 1$, $D^{ps} = -1$, $C_{ps} = 0$, $V_0 = -0.2$, $V_1 = 0.1$.

*SS*: Energy vs. $\alpha$ for Spin Symmetry Limit for second choice
$H = 0.5$, M=5fm$^{-1}$, $A^s = 1$, $B^s = -2$, $C^s = 1$, $D^s = -1$, $C_s = 0$, $V_0 = 0.2$, $V_1 = 0.1$.

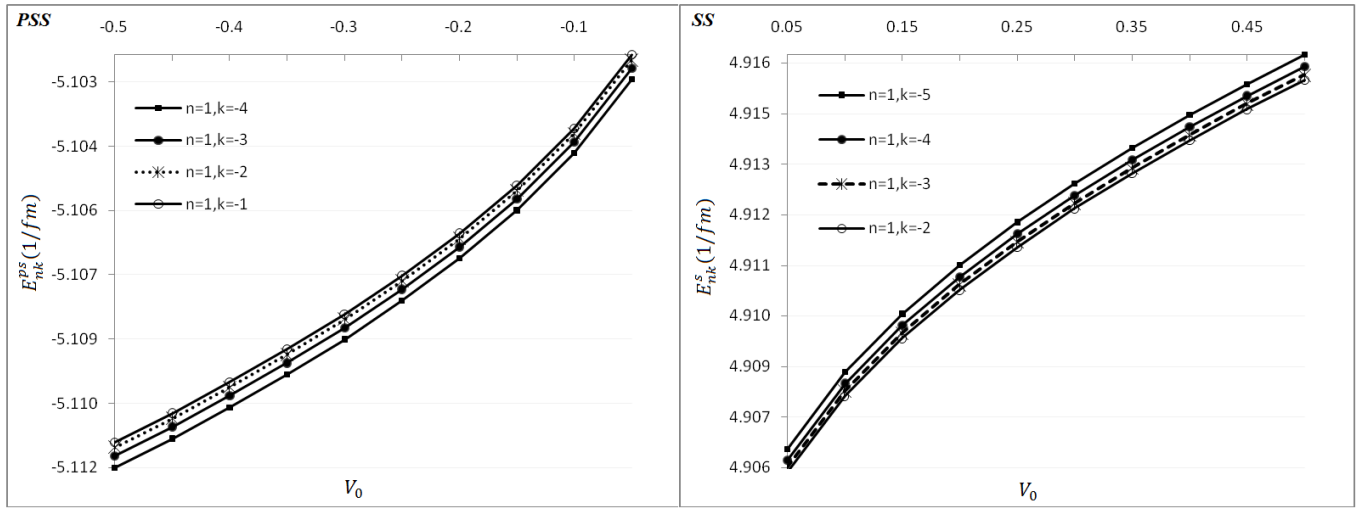

**Fig. (7).** *PSS*: Energy vs. $V_0$ for Pseudospin Symmetry Limit for second choice
$H = 0.5$, $\alpha = 0.01$ , M=5fm$^{-1}$, $A^{ps} = 1$, $B^{ps} = -2$, $C^{ps} = 1$, $D^{ps} = -1$, $C_{ps} = 0$, $V_1 = 0.1$.

*SS*: Energy vs. $V_0$ for Spin Symmetry Limit for second choice
$H = 0.5$, $\alpha = 0.01$ , M=5fm$^{-1}$, $A^s = 1$, $B^s = -2$, $C^s = 1$, $D^s = -1$, $C_s = 0$, $V_1 = 0.1$.

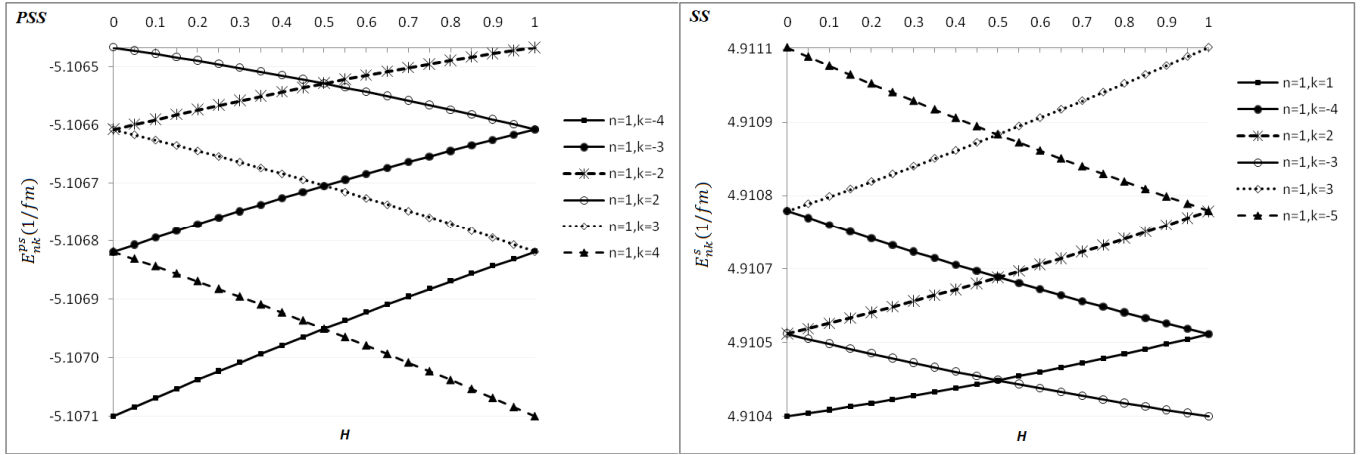

**Fig. (8).** *PSS*: Energy vs. $H$ for Pseudospin Symmetry Limit for second choice
$\alpha$=0.01 , M=5fm$^{-1}$, $A^{ps}=1$, $B^{ps}=-2$, $C^{ps}=1$, $D^{ps}=-1$, $C_{ps}=0$, $V_0=-0.2$, $V_1=0.1$.

*SS*: Energy vs. $H$ for Spin Symmetry Limit for second choice
$\alpha$=0.01 , M=5fm$^{-1}$, $A^{s}=1$, $B^{s}=-2$, $C^{s}=1$, $D^{s}=-1$, $C_s=0$, $V_0=0.2$, $V_1=0.1$.

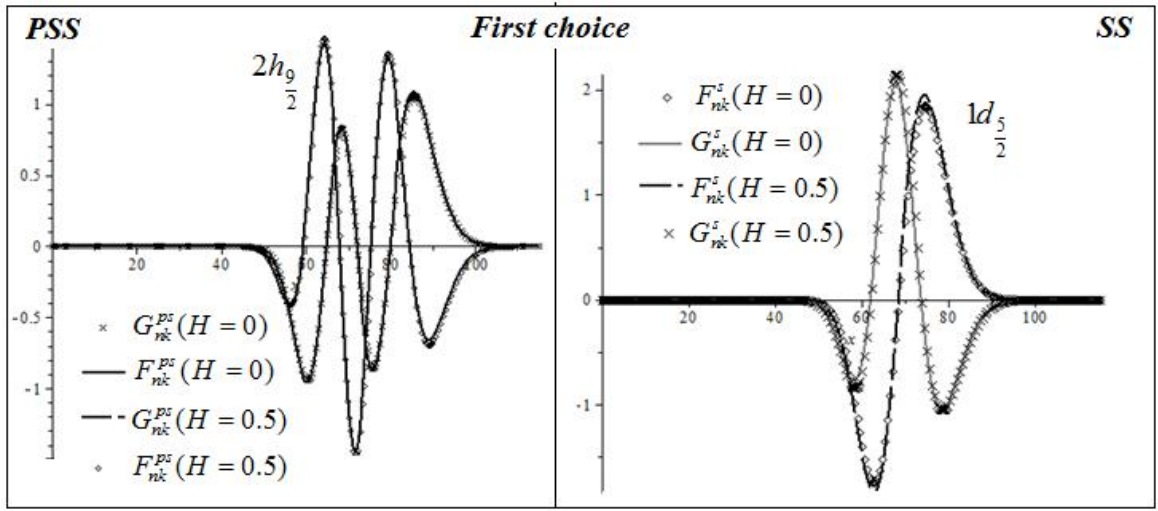

**Fig. (9).** *PSS*: Lower and upper radial wave functions in view of the pseudospin symmetry for
$H=0.5, \alpha$=0.01 , M=5fm$^{-1}$, $A^{ps}=1$, $B^{ps}=-2$, $C^{ps}=1$, $D^{ps}=-1$, $C_{ps}=0$, $V_0=-0.2$, $V_1=0.1$.

*SS*: Lower and upper radial wave functions in view of the spin symmetry for
$H=0.5, \alpha$=0.01 , M=5fm$^{-1}$, $A^{s}=1$, $B^{s}=-2$, $C^{s}=1$, $D^{s}=-1$, $C_s=0$, $V_0=0.2$, $V_1=0.1$.

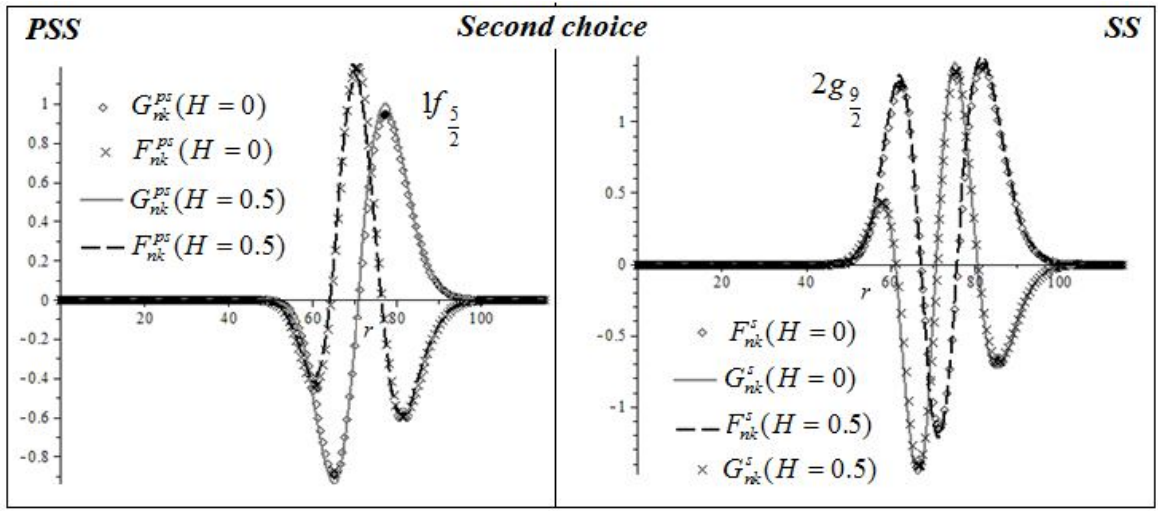

**Fig. (10). *PSS***: Lower and upper radial wave functions in view of the pseudospin symmetry for $H = 0.5, \alpha = 0.01$ , M=5fm$^{-1}$, $A^{ps} = 1$, $B^{ps} = -2$, $C^{ps} = 1$, $D^{ps} = -1$, $C_{ps} = 0$, $V_0 = -0.2$, $V_1 = 0.1$.

***SS*:** Lower and upper radial wave functions in view of the spin symmetry for $H = 0.5, \alpha = 0.01$ , M=5fm$^{-1}$, $A^s = 1$, $B^s = -2$, $C^s = 1$, $D^s = -1$, $C_s = 0$, $V_0 = 0.2$, $V_1 = 0.1$.